\documentclass[aps,prc,amsmath,amssymb,twocolumn,showpacs]{revtex4-1} 
\usepackage{graphicx,amsmath,amssymb,bm,soul,physics,comment}
\usepackage[section]{placeins}
\usepackage[caption=false]{subfig}	

\usepackage[usenames]{color}

\newcommand{\bvec}[1]{\mathbf{#1}}
\newcommand{\bvecg}[1]{\boldsymbol{#1}}

\begin{document} 
 
 \title{Skyrme-based extrapolation for the static response of neutron matter}

\author{Mateusz Buraczynski, Samuel Martinello, and Alexandros Gezerlis}
\affiliation{Department of Physics, University of Guelph, 
Guelph, ON N1G 2W1, Canada}

\begin{abstract} 
The study of inhomogeneous neutron matter can provide insights into 
the structure of neutron stars as well as their dynamics in neutron-star mergers. 
In this work we tackle pure neutron matter in the presence of a periodic external
field by considering 
a finite (but potentially large) number of particles placed in periodic boundary conditions. 
We start with the simpler setting of a noninteracting gas and then switch to a
Skyrme-Hartree-Fock approach, showing static-response results for five distinct
Skyrme parametrizations. 
We explain both the technical details of our computational approach, as well as 
the significance of these results as a general finite-size extrapolation scheme
that may be used by \textit{ab initio} practitioners to approach the static-response
problem of neutron matter in the thermodynamic limit. 
\end{abstract} 
\maketitle 

\section{Introduction}

The examination of neutron matter properties provides an important basis for the understanding of the structure and dynamics of neutron stars. Pure neutron matter
constitutes a strongly interacting many-body system which has connections
to the nuclear symmetry energy, as well as the beta-stable matter which exists in
the interior of 
neutron stars. Fascinating connections exist between the microphysics of 
neutron matter and the study of gravitational waves: we are now in the new
era of multimessenger astronomy, during which the merging of two neutron
stars can be directly detected. Finally, neutron matter at low density
is quite similar to ultracold atomic systems probed in a number of laboratories
around the world~\cite{Gandolfi:2015,Lattimer:2016,Abbott:2017,Tews:2018,Tong:2020,Gezerlis:2008, Boulet:2018}.

In recent decades, considerable progress has been made in tackling 
neutron matter using \textit{ab initio} many-body techniques, namely
first-principles computational methods which directly solve the 
many-body Schr\"odinger equation using controlled approximations. 
Such approaches typically produce as one output the equation of state (EOS)
of neutron matter~\cite{Friedman:1981,Akmal:1998,Schwenk:2005,Epelbaum:2009,Kaiser:2012,Friman:2019}. A lot of effort has been expended in a related direction,
namely that of producing high-quality microscopic two- and three-nucleon 
interactions~\cite{Carlson:Morales:2003,Gandolfi:2009,Gezerlis:2010,Gandolfi:2012,Baldo:2012,Hebeler:2010,Tews:2013,Gezerlis:2013,Coraggio:2013,Hagen:2014,Gezerlis:2014,Carbone:2014,Roggero:2014,Wlazlowski:2014,Tews:2016,Soma:2020}.
A distinct approach is more phenomenological: energy-density functional
theories for nuclei and for infinite matter 
(sometimes cast in the form of a Hamiltonian formalism with an effective interaction) employ a number of parameters which are fit to observed properties of nuclei and 
matter, as well as to ``synthetic data'' coming from \textit{ab initio} 
computations~\cite{Bender:2003,Chabanat:1998,Rios:2014,Alam:2014,Basilico:2015,Lacroix:2016,Chabanat:1998, Fayans:1998,Brown:2000,Chappert:2008,Fattoyev:2010,Fattoyev:2012,Brown:2014,Rrapaj:2016,Chamel:2008,Forbes:2014,Roggero:2015,Pudliner:1996,Pederiva:2004,Gandolfi:2011,Potter:2014,Gandolfi:2016}.

The study of neutron matter is not limited to ground-state properties.
For example, the single-particle excitation spectrum (and the related concept
of the effective mass) are related to a neutron star's maximum mass as well
as to giant quadrupole resonances~\cite{Li:2018,Isaule:2016,Grasso:2018,Bonnard:2018,Buraczynski:2019,Buraczynski:2020,Bonnard:2020}. The present paper revolves
around another way of going beyond the ground state, namely the placing
of strongly interacting matter within a periodic external potential. This latter
problem is known as that of the ``static response'' of neutron matter:
in addition to being important in and of its own, it is also related to 
optical lattices in cold-atom physics as well as to the study of neutron star
crust (which involves both nuclei and extended neutron matter, which is therefore
inhomogeneous). 

Most work on the static response of neutron matter has been in the context
of a mean-field-like technique~\cite{Pastore:2015,Iwamoto:1982,Olsson:2004,Chamel:2012,Chamel:2013,Kobyakov:2013,Pastore:2014,Chamel:2014,Davesne:2015}.
\textit{Ab initio} work has been limited to quantum Monte Carlo (QMC) studies~\cite{Buraczynski:2016,Buraczynski:2017,Buraczynski:2021},
but there is no \textit{a priori} obstacle keeping other techniques from tackling
this problem in the future.
Most such first-principles techniques are forced to use a finite particle number
due to computational-scaling considerations. As will become clearer in what follows,
the finite-size (FS) effects resulting from the use of such a finite (typically
small) particle number are much more pronounced in the study of inhomogeneous
matter than they are for homogeneous matter. Thus, it is worthwhile 
to produce a general-purpose finite-size
extrapolation scheme that makes use of a more widely applicable technique;
this was reported on in Ref.~\cite{Buraczynski:2021}, without discussing implementation
aspects or showing intermediate results.
The present paper discusses the formalism and results in considerable detail, so
that other practitioners may be able to benefit from them in the future.

\section{Static-response theory}
\label{sec:static}

Consider a homogeneous and isotropic system such as neutron matter. This may be described by a Hamiltonian $\hat{H}_0$ with the ground state $\Psi_0(\bvec{r})$, energy $E_0$ and number density $\rho_0(\bvec{r}) = |\Psi_0(\bvec{r})|^2$. The constant density is denoted by $\rho_0(\bvec{r}) = \rho_0$. With the addition of a static external potential $v(\bvec{r})$, the Hamiltonian for this perturbed system may now be written as:
\begin{align}
	\hat{H}_v = \hat{H}_0 + \int d\bvec{r}\hat{\rho}(\bvec{r})v(\bvec{r})
	\label{eq:perturbation}
\end{align}
where effects due to $v(\bvec{r})$ are coupled through the one-body density operator $\hat{\rho}(\bvec{r}) = \sum_{i=1}^N\delta(\bvec{r}-\bvec{r}_i)$. $N$ is the number of particles in a finite volume $V$ satisfying $N/V = \rho_0$ which is maintained in the thermodynamic limit $(V\rightarrow \infty, N\rightarrow \infty)$. With these definitions in place, we see that the perturbed Hamiltonian is a functional of the external potential $v(\bvec{r})$ and we may in turn write the ground state energy $E_v = E_0[v]$ and density $\rho_v(\bvec{r}) = \rho_0(\bvec{r}, [v])$ as functionals of this potential. The functional expansion for the density $\rho_v(\bvec{r})$ with respect to $v(\bvec{r})$ is:
\begin{align}
	&\rho_v(\bvec{r}) 
	=\rho_0+ \nonumber\\
	& \sum_{k=0}^{\infty} \frac{1}{k!}\int d\bvec{r}_1 \cdots d\bvec{r}_k \chi^{(k)} (\bvec{r}_1 - \bvec{r}, \dots , \bvec{r}_k - \bvec{r})v(\bvec{r}_1)\cdots v(\bvec{r}_k)
	\label{eq:densityexpansion}
\end{align}	
where we note that $\rho_0(\bvec{r},[0]) = \rho_0$ is the unperturbed uniform density and the 
\begin{align}
	\chi^{(k)} (\bvec{r}_1 - \bvec{r}, \dots , \bvec{r}_k - \bvec{r}) = \frac{\delta^k \rho_0(\bvec{r},[v])}{\delta v(\bvec{r}_1)\cdots\delta v(\bvec{r}_k)}\Bigg|_{v=0}
	\label{eq:responsevariation}
\end{align}
terms are the (static) density-density response functions.

We now write an analogous expression for the energy $E_v$ using first-order perturbation theory to derive the appropriate functional form: 
\begin{align}
	&E_v
	= E_0
	+ \rho_0\int d\bvec{r} v(\bvec{r})
	+ \sum_{k=1}^{\infty} \frac{1}{(k+1)!} \times \nonumber\\&
	\int d\bvec{r} d\bvec{r}_1 \cdots d\bvec{r}_k \chi^{(k)} (\bvec{r}_1 - \bvec{r}, \dots , \bvec{r}_k - \bvec{r})v(\bvec{r})v(\bvec{r}_1)\cdots v(\bvec{r}_k)
	\label{eq:perturbenergyfunctionalchi}
\end{align}
The external potential $v(\mathbf{r})$ may be decomposed into its Fourier components:
\begin{align}
	v(\bvec{r}) = \sum_\bvec{q} v_\bvec{q} \exp[i\bvec{q}\cdot\bvec{r}]
	\label{eq:potentialfourier}
\end{align}
Both Eq.~(\ref{eq:responsevariation}) and Eq.~(\ref{eq:perturbenergyfunctionalchi}) can then be rewritten to express the variation in density and energy per particle in wave-number space:
\begin{align}
	&\delta\rho(\bvec{r}) \equiv \rho_v(\bvec{r}) - \rho_0 
	= \sum_{k=1}^{\infty}\frac{1}{k!} 
	\sum_{\bvec{q}_1,\dots,\bvec{q}_k}\chi^{(k)}(\bvec{q}_1,\dots,\bvec{q}_k) \times\nonumber\\
	&v_{\bvec{q}_1}\cdots v_{\bvec{q}_k}\exp[i(\bvec{q}_1+\cdots+\bvec{q}_k)\cdot\bvec{r}]
	\label{eq:densityfourier}
\end{align}
and
\begin{align}
	&\delta\bar{E}(\bvec{r}) \equiv \frac{E_v}{N} - \frac{E_0}{N} 
	=v_0+\nonumber\\
	 &\frac{1}{\rho_0}\sum_{k=1}^{\infty}\frac{1}{(k+1)!}
	\sum_{\bvec{q}+\bvec{q}_1+\cdots+\bvec{q}_k=0}\chi^{(k)}(\bvec{q}_1,\dots,\bvec{q}_k)v_{\bvec{q}}v_{\bvec{q}_1}\cdots v_{\bvec{q}_k}
	\label{eq:energyfourier}
\end{align}
where the bar notation means energy per particle.
For a monochromatic potential of the form $v(\bvec{r}) = 2v_\bvec{q}\cos(\bvec{q}\cdot\bvec{r}) = v_\bvec{q}(\exp[i\bvec{q}\cdot\bvec{r}] + \exp[-i\bvec{q}\cdot\bvec{r}])$, Eq.~(\ref{eq:densityfourier}) and Eq.~(\ref{eq:energyfourier}) 
to 3$^{rd}$ and 4$^{th}$ order in $v_\bvec{q}$, respectively, take the form:
\begin{align}
	\delta\rho(q,z) 
	&= 
	2\cos(qz)\chi^{(1)}(q)v_q \nonumber\\&
	+ \cos(qz)\chi^{(3)}(q,q, -q)v_q^3
	+ \mathcal{O}(v_q^5)
	\label{eq:densityvariation}
\end{align}
and
\begin{align}
	\delta\bar{E}(q) 
	&= 
	\frac{1}{\rho_0}\chi^{(1)}(q)v_q^2
	+ \frac{1}{4\rho_0}\chi^{(3)}(q,q, -q) v_q^4
	+ \mathcal{O}(v_q^6) 
	\label{eq:energyvariation}
\end{align}
where we took, without sacrificing generality, our coordinate system's $z$ axis
to point along $\mathbf{q}$ (thereby giving rise to $\cos(qz)$).

For fixed values of $\bvec{q}$ (the periodicity of the potential), the energy per particle of the system can be evaluated at various $v_q$ values (external potential strength parameter). The polynomial described in Eq.~(\ref{eq:energyvariation}) is then used to produce an even-degree polynomial fit in powers of $v_\bvec{q}$ to these points. We are interested in the leading term of this fit, $\frac{1}{\rho_0}\chi^{(1)}(q)v_q^2$, from which the linear density-density response function $\chi^{(1)}(q)$ may be extracted. It is important to note that the fidelity of this fit is dependent on both the number of external potential strengths sampled as well as the order of the polynomial fit used, with fits of degree higher than two producing better results in comparison to a simple quadratic fit. We have found that for the number of points used in our fits, a fourth degree fit provides the best results while mitigating inaccuracies due to over-fitting. More details on the fit are given in later sections.

The dynamical response function of a system is related to its dynamical structure factor (as per the fluctuation-dissipation theorem). The linear static-response function can be related to this as well via the Kramers-Kronig relations. At zero temperature the relation is:

\begin{align}
\chi(q)=-\frac{\rho_0}{\pi\hbar}\int_0^{\infty}d\omega \frac{S(q,\omega)}{\omega}
\end{align}
where $S(q,\omega)$ is the dynamical structure factor.
In the $q\to 0$ limit this expression is related to the thermodynamic compressibility as:
\begin{align}
\frac{1}{\chi(0)}&=-\frac{1}{\rho_0}\Bigg (\frac{\partial p}{\partial \rho_0}\Bigg	 )_{T=0}
\label{eq:comp}
\end{align}
This is known as the compressibility sum rule. We can express this in terms of the EOS of the homogeneous system. At zero temperature there is no heat transfer so pressure is given by:
\begin{align}
p&=-\Bigg(\frac{\partial E}{\partial V}\Bigg)_{T=0} =-\Bigg(\frac{\partial E}{\partial \rho_0}\Bigg)_{T=0}\Bigg(\frac{\partial \rho_0}{\partial V}\Bigg)_{N}\nonumber\\
&=\frac{N}{V^2}\Bigg(\frac{\partial E}{\partial \rho_0}\Bigg)_{T=0}
=\rho_0^2\Bigg(\frac{\partial \bar{E}}{\partial \rho_0}\Bigg)_{T=0}
\label{eq:pressure}
\end{align}
where the second step is an application of the chain rule and the third step is from $\rho_0=N/V$.
Rearranging Eq.~(\ref{eq:comp}) at zero temperature and using Eq.~(\ref{eq:pressure}):
\begin{align}
\frac{1}{\chi(0)}&=-\frac{1}{\rho_0}\Bigg (\frac{\partial p}{\partial \rho_0}\Bigg	 )_{T=0}\nonumber\\
&=-\Bigg[2\Bigg(\frac{\partial \bar{E}}{\partial \rho_0}\Bigg)_{T=0}+\rho_0\Bigg(\frac{\partial^2 \bar{E}}{\partial \rho_0^2}\Bigg)_{T=0}\Bigg]\nonumber\\
&=-\frac{\partial^2(\rho_0\bar{E})}{\partial \rho_0^2}\Bigg|_{T=0}
\label{eq:comp2}
\end{align}
Thus the EOS constrains the $q=0$ response function.

\section{Non-interacting problem}
\label{sec:finite}

As a preliminary step before tackling the interacting neutron matter problem, we first characterize the properties of an infinite system of a free, non-interacting Fermi gas (FFG) in three dimensions. Not only does this system provide formalism and guidance in tackling the interacting problem, we also use the free-gas as a first step towards applying FS fixes to the response function computed by one's finite many-body method of choice.

In free space, the single-particle spatial wavefunctions $\phi_i(\bvec{r})$ are governed by the independent-particle time-independent Schr{\"o}dinger equation:
\begin{align}
	-\frac{\hbar^2}{2m}\nabla^2\phi_i(\bvec{r}) = e_i\phi_i(\bvec{r}) \label{eq:freehamiltonian3d}
\end{align}
where $e_i$ is the eigenenergy associated to the state $\phi_i(\bvec{r})$. The set of solutions are plane-waves which are non-normalizable. The standard method for dealing with this comes from considering translational invariance. We consider $N$ particles contained in a box of volume $V=L^3$ and apply periodic boundary conditions to this box. The wave-function is normalized within this volume. The standard is to center the box at the origin and have the box extend from $-L/2$ to $L/2$ in all three Cartesian dimensions. The periodic boundary conditions are that the value of the wave-function does not change when shifting any of the $x,y,z$ coordinates by an integer multiple of $L$. The infinite system is considered to be an infinite tessellation of these boxes due to the translational invariance. The assumption that in the thermodynamic limit which takes $N\rightarrow\infty,\: V\rightarrow\infty: N/V =$~constant, the properties of the system will have converged to their thermodynamic values. 

The single-particle wave function solutions normalized to the box are then given by: $\phi_i(\bvec{r}) = (1/\sqrt{V})e^{i\bvec{k}\cdot\bvec{r}}$ with single-particle energies $e_i = \hbar^2\bvec{k}^2/2m$ and the periodic boundary conditions impose the restriction $\bvec{k} = (2\pi/L)(n_x,n_y,n_z)$ for $n_x,n_y,n_z \in \mathbb{Z}$.
For a Fermi gas, the lowest-energy states are filled first according to the Pauli exclusion principle. In the case of neutrons, a spin-1/2 system, each state may be populated by each of the +1/2 and -1/2 spin projections. With this in place, 
the energy of the finite system may be simply calculated as the sum of the 
single-particle energies: 
\begin{align}
	E_{FG} = \frac{\hbar^2}{2m}\sum_{|\bvec{k}|<|\bvec{k}|_{max},\sigma}\bvec{k}^2 
\end{align}
In the TL, the occupied k-space becomes a sphere and the maximal occupied wave-vector magnitude $k_F$ is the radius of this sphere. The relationship to number density is $k_F = (3\pi^2\rho_0)^{1/3}$. The energy per particle in the TL is:
\begin{align}
	\bar{E}_{FG} = \frac{3}{5}E_F = \frac{3}{5}\frac{\hbar^2k_F^2}{2m}
\end{align}
In the context of many-body methods applied to complex systems, calculations will often be limited to a finite number of particles restricted to a box similarly to the finite treatment of the non-interacting problem. This results in discrepancies from the TL which are called finite size (FS) effects. The nice thing about studying simpler systems like the non-interacting Fermi gas is that one is also able to perform calculations for large particle numbers, as well as in the TL directly. The discrepancies between small and large particle number results for the free gas gives insight into FS effects. One may choose to apply these to the complex system for the purpose of extrapolating finite-particle results to the TL. 

Since we are forced to deal with FS effects, it is useful to try and minimize these prior to extrapolating to the TL. This can be handled in part via more complicated boundary conditions but it is also important to choose an appropriate number of particles. Typically, one works at shell-closures of the free-Fermi gas where the energy per particle is close to the TL value. A shell closure is defined as any particle number where all available states at the highest occupied energy level have been occupied. Shell closures are meant to remove the ambiguity of state selection as much as possible; this ambiguity may be unavoidable where one introduces an external perturbation such as in the static-response problem. The particle number closures will vary at larger strengths of the external potential. We work with a free-gas particle number closure and limit the size of the external strength parameter to minimize shell-effects. Note that the relevance of free-gas shell closure to a many-body method is clear in methods such as Quantum Monte Carlo where the trial wave-function is built from the non-interacting many-body wavefunction.

\begin{figure}[t]
\begin{center}
\includegraphics[width=1.0\columnwidth,clip=]{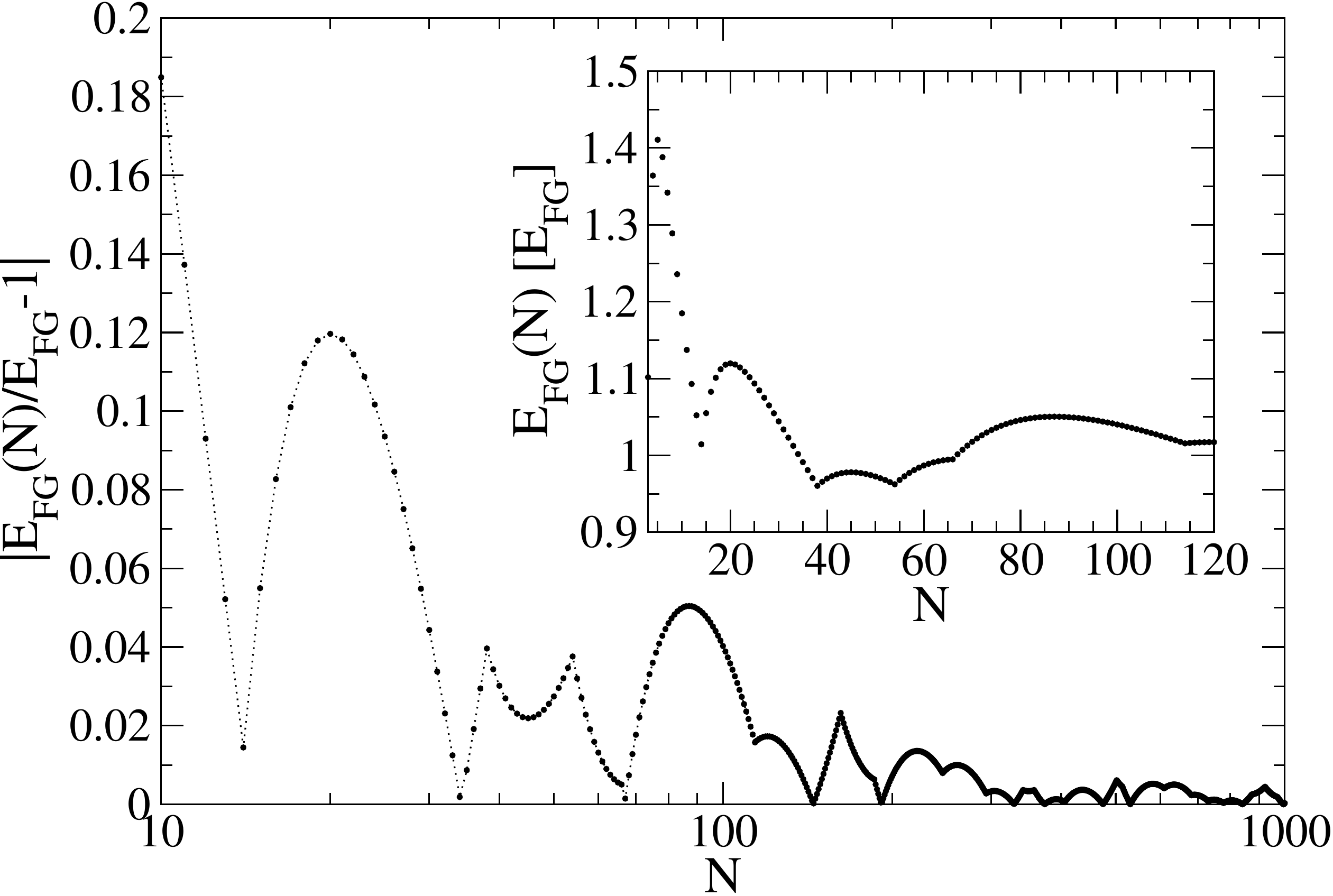}
\caption[FFG finite size effects]{Relative error of $N$ particle energy compared to TL energies shows magnitude of finite size effects versus particle number. 
Inset: energy FS effects on a linear scale. \label{fig:FFG-TL}}
\end{center}
\end{figure}

The shell-closures for the free-Fermi gas occur at $N=\{2,14,38,54,66,114...\}$. For a many-body method such as Quantum Monte Carlo applied to nuclear systems, the computational complexity becomes unmanageable somewhere above 100 particles. It is fortuitous that there exists a local minimum in relative error from the TL energy per particle at $67$ particles. The relative error is still quite small at $66$ particles, one of the shell closures. In fact it is even smaller than the error at $114$ particles. This can be seen in Fig.~\ref{fig:FFG-TL} where the relative error is plotted versus a logarithmic scale in the particle number. For this reason, $66$ particles are widely used by \textit{ab initio} many-body methods in studies of neutron matter. We have done the same and have therefore also used $66$ particles in our noninteracting-gas calculations.

We now turn to the static response of the noninteracting Fermi gas. This is our starting point for illustrating the steps of extracting the response function from energy calculations. In a research problem, one would apply these steps to the results obtained from some many-body method. More specifically, in this paper we show how one may extrapolate their energy results and then apply this methodology to compute response in the TL. Although the extrapolation is done on the energy per particle level, computing the response functions of simpler systems at both small and large particle numbers gives an appropriate qualitative look at the size of FS effects. With all this in mind, we limit this section to response calculations in the noninteracting Fermi gas.

As described in the response section, consider a perturbation of the form $v(\bvec{r}) = 2v_q\cos(qz)$. $2v_q$ is the amplitude of the potential. We have decided to always work with a whole number of periods of the potential in the box of length $L$: i.e. $q= 2\pi n/L$, where $n$ is this whole number. This is done to make the calculations easier and to preserve the translational invariance. The single particle free-gas Schr{\"o}dinger equation is modified to:
\begin{align}
	-\frac{\hbar^2}{2m}\nabla^2\phi_i(\bvec{r})+  2v_q\cos(qz)\phi_i(\bvec{r}) = e_i\phi_i(\bvec{r})
\label{eq:modifiedSchrodinger} 
\end{align}
The solutions to this equation have been modified in the z-component. That is, after separation of variables so that $\phi_i({\bvec r})=X_i(x)Y_i(y)Z_i(z)$, the orbitals in x and y remain as plane-waves: $X(x)=[1/\sqrt(L)]e^{ik_xx}$, $Y(y)=[1/\sqrt(L)]e^{ik_yy}$. The orbitals in z are the well studied Mathieu functions \cite{NIST:DLMF}. The energy contribution from a single-particle state has been separated into x,y, and z contributions from their respective orbitals. The total energy and many-body wave-function are given by filling up the lowest available single-particle states analogously to the free-Fermi gas problem. Alternatively, the calculations may be performed via solving the Hartree-Fock equations as described later and will yield consistent results.

The values of $q$ at which we can compute $\chi(q)$ are limited by the restriction described above. We mainly studied $q$ values corresponding to $1, 2, 3, 4, 6, 8,\, {\rm and}\, 10$ periods of the potential traversing the box containing $66$ particles. Thus, we have extracted $\chi(q)$ at the lowest $4$ available $q$ values, and have skipped some of the larger available values which are less interesting. 

The steps to computing $\chi(q)$ are as follows: Choose a density and a particle number to work at. These are fixed for all of the calculations. The box-size is determined by these and the available values of $q$ are restricted in turn. Select a value of $q$ to compute the response at. Compute the energy per particle of the system with external potential $2v_q\cos(qz)$ for several values of the strength parameter $v_q$. After obtaining a set of these values which we label $\bar{E}(v_q)$, subtract out the non-perturbed energy $\bar{E}_{0}(N)$ so that $\Delta\bar{E}(v_q)=\bar{E}(v_q)-\bar{E}_{0}(N)$. Perform a least-squares fit to a truncated version of Eq.~(\ref{eq:energyvariation}):
\begin{align}
\Delta\bar{E}\approx\frac{\chi^{(1)}(q)}{\rho_0}v_q^2+C_4v_q^4
\label{eq:fitformEq}
\end{align}
and $\chi(q)/\rho_0$ is extracted as the coefficient to the quadratic component of this quartic.

We have selected the values of $v_q$ with Quantum Monte Carlo simulations in mind. This is because our FS prescription requires one to compute the energy per particle in both the complex system and the simpler system at the same $q$ and $v_q$ values. As a result we have excluded very small values of $v_q$ for which the error in QMC would dominate over the slight change in energy induced by the perturbation. We also do not go to very large values of $v_q$ in order to minimize shell effects (as mentioned earlier) as well as to keep the truncation of Eq.~(\ref{eq:energyvariation}) as valid as possible. The values we use are $2v_q/E_F=0.25, 0.3, 0.35, 0.5$.

We need to compute $\chi(q)$ for both small and large particle numbers. For the purposes of our FS prescription, we must compute energy per particles at the exact same set of $q$ and $v_q$ values for both particle numbers. Typically $66$ is the smaller particle number. The choice of density has already restricted the box-size and the available periodicities for the potential. In order to preserve a whole number of periods in the box, we only work with boxes with linear dimensions that are integer multiples of the small box containing $66$ particles. Thus in in moving towards the TL starting from $N=66$ we are restricted to $N=\{66,528,1782,4224,8250,...\}$. 
We now introduce the first version of our FS prescription which estimates FS effects from the noninteracting Fermi gas:
\begin{align}
&\Delta\bar{E}(TL)=\nonumber\\
&\Delta\bar{E}_{\rm{ab~initio}}(N_{\rm{small}})-\Delta\bar{E}_{NI}(N_{\rm{small}})+\Delta\bar{E}_{NI}(N_{\rm{big}})
\label{eq:prescriptionfree}
\end{align}
where the \textit{ab initio} subscript refers to any first-principles many-body method. The $NI$ subscript is for the non-interacting gas. It is implied that the changes in energy per particle are all evaluated at the same $q$ and $v_q$ value. The prescription applies this equation to every $q$ and $v_q$ case. Thus in words, we have estimated that the FS effects may be removed by shifting the change in energy per particle by the same amount as it shifts in the noninteracting Fermi gas when moving from small to large particle numbers. We have also taken $N_{\rm{small}}=66$ and $N_{\rm{big}}=8250$ for this work. The TL response is computed by applying the fitting method described earlier to the extrapolated value in the LHS of Eq.~(\ref{eq:prescriptionfree}).

\begin{table}[t]
\begin{center}
\caption[Finite size error: 66 vs. 8250 particles]{Relative error of the linear response function due to finite size effects for 66 and 8250 particles at a density of $0.10$ fm$^{-3}$}
\label{table:66vs8250error}
\begin{tabular}{ccc}
	\hline
	\hline
	$q/q_F$  & $N=66$ Error [$\%$]  & $N=8250$ Error [$\%$]\\
	\hline
	0.503 & 3.26 & 1.67 \\
1.005 & 29.3 & 1.03 \\
1.508 & 2.96 & 0.879 \\
2.010 & 18.3 & 0.785 \\
3.015 & 0.575 & 0.0261 \\
4.021 & 0.0917 & 0.0217 \\
5.026 & 0.0231 & 0.0149 \\
	\hline
	\hline
\end{tabular}
\end{center} 
\end{table}

\begin{figure}[t]
\begin{center}
\includegraphics[width=1.0\columnwidth,clip=]{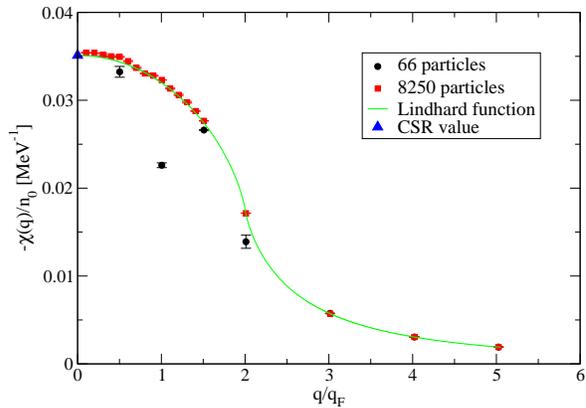}
\caption{Noninteracting Fermi gas linear response functions at a density of $0.1\,{\rm fm}^{-3}$. The circles and squares correspond to 66 and 8250 particles respectively. The line is the TL response function.}\label{fig:free0.1resp}
\end{center}
\end{figure}

We determined that $8250$ particles sufficiently approximates the TL for our purposes. This can be seen in Fig.~\ref{fig:free0.1resp} where the non-interacting response at a density of $0.1\,{\rm fm}^{-3}$ is estimated for both $66$ (circles) and $8250$ (squares) particles. The CSR value is shown at $q=0$ (triangle). The solid curve is the TL response for the noninteracting gas known as the Lindhard function:
\begin{align}
	\chi_L(q) = - \frac{mq_F}{2\pi^2\hbar^2}\Bigg[1+\frac{q_F}{q}\Bigg(1-\Big(\frac{q}{2q_F}\Big)^2\Bigg)\ln\Bigg|\frac{q+2q_F}{q-2q_F}\Bigg|\Bigg]
\end{align}
The $8250$ particle response agrees well with the Lindhard function across all $q$ values. There are some discrepancies at low $q$ which are much too small to impact our results. The relative error which is always below $2\%$ is 
small in comparison to other errors typically present in quantum many-body theory. It is also apparent that both finite particle responses agree well with the TL for $q/q_F \geq 3$. However, unlike the $8250$ case, the $66$ particle response displays large deviations from the TL at lower $q$ values.  We can see in Table~\ref{table:66vs8250error} that the relative errors in the 66 particle case have a maximum value of 29.3\% at $q/q_F \approx 1$. These deviations are the FS effects and it is at precisely these $q$ values where the FS prescription has the largest impact on response extrapolation to the TL. The noninteracting response has been estimated for many more $q$ values for $8250$ particles compared to $66$ particles. These $q$ values are not available for the smaller system because the periodicities do not respect the translational invariance. For the remainder of this work, we only present responses at the same set of $q$ values for both $66$ and $8250$ particles.

We wrap up this section with an example of the compressibility sum rule. We apply Eq.~(\ref{eq:comp2}) to the noninteracting Fermi gas:
\begin{align}
\chi(0)&=-\Bigg[\frac{\partial^2(\rho_0\bar{E}(\rho_0))}{\partial \rho_0^2}\Bigg]^{-1}\nonumber\\
&=-\Bigg[\frac{\partial^2}{\partial \rho_0^2}\Bigg(\frac{3}{5}\frac{\hbar^2}{2m}(3\pi^2)^{2/3}\rho_0^{5/3}\Bigg)\Bigg]^{-1}
&=-\frac{mq_F}{\pi^2\hbar^2}
\end{align}
which agrees with the $q \to 0$ limit of the Lindhard function as can be seen in Fig.~\ref{fig:free0.1resp}.

We applied Eq.~(\ref{eq:prescriptionfree}) to QMC calculations of neutron matter in previous work~\cite{Buraczynski:2016,Buraczynski:2017} but found large discrepancies between the low-$q$ response and the CSR values. We have since published results using an improved prescription which goes beyond the noninteracting Fermi gas. 
In Ref.~\cite{Buraczynski:2021} we reported on an improved extrapolation scheme
based on Skyrme-Hartree-Fock calculations; it is to the details of this approach
that we now turn.

\section{Skyrme energy-density functionals}

The limitations in the prescription thus far can be attributed to the lack of information concerning the inter-particle interactions in the extrapolation. It is reasonable to consider a phenomenological theory in place of the non-interacting gas, but one which is still able to handle many more particles than 
would be possible in a first-principles calculation. The Skyrme interaction was specifically designed to provide a framework from which one could study nuclear matter without many of the limitations imposed by the full problem. Thus, while one loses
contact with the microscopic many-body Schr\"odinger equation, the advantage 
that a new class of further calculations now becomes accessible (which are not going
to be carried out any time soon using an \textit{ab initio} technique).

Skyrme's interaction approximates the nuclear potential through the composition of two- and three-body effective interaction terms $v^{(2)}_{ij}$ and $v^{(3)}_{ijk}$:
\begin{align}
	V = \sum_{i<j} v^{(2)}_{ij} +\sum_{i<j<k} v^{(3)}_{ijk}
\end{align}
The two-body interaction may be written as a short-range expansion in coordinate space of the form \cite{Bender:2003,Skyrme:1958,SLy4-1}:
\begin{align}
	v^{(2)}_{ij} 
	&= t_0(1+x_0P_\sigma)\delta(\bvec{r}_i-\bvec{r}_j) \nonumber\\
	&+\tfrac{1}{2}t_1(1+x_1P_\sigma)[\delta(\bvec{r}_i-\bvec{r}_j)\bvec{P}^2+\bvec{P}^{\prime 2}\delta(\bvec{r}_i-\bvec{r}_j)] \nonumber\\
	&+ t_2(1+x_2P_\sigma) \bvec{P}^\prime \cdot \delta(\bvec{r}_i-\bvec{r}_j) \bvec{P}
\end{align}
where $P_\sigma$ is a spin-exchange operator. The $t_s, x_s : s=1,2,3$ and $t_0$ parametrize the mean central potential. All are free parameters corresponding to the constraints of the system. Additionally, the $\bvec{P}$ is the operator $(\bvec{\nabla}_1-\bvec{\nabla}_2)/2i$ acting to the right and $\bvec{P}^\prime$ is the operator $-(\bvec{\nabla}_1-\bvec{\nabla}_2)/2i$ acting to the left. The three-body interaction potential can similarly be taken to be a zero-range interaction of the form:
\begin{align}
	v^{(3)}_{ijk} 
	= t_3\delta(\bvec{r}_i-\bvec{r}_j)\delta(\bvec{r}_j-\bvec{r}_k) \nonumber\\
\end{align}
or, more commonly:
\begin{align}
	v^{(3)}_{ijk} 
	= \frac{1}{6}t_3(1+x_3P_\sigma)\delta(\bvec{r}_i-\bvec{r}_j)\rho\Big(\frac{\bvec{r}_i+\bvec{r}_j}{2}\Big)^\alpha
\end{align}
which is parametrized by the density $\rho$ and the parameters $t_3$, $x_3$ and $\alpha$. It is important to notice that both the two-body and three-body interactions are each written in terms of $\delta$-distributions as this provides an easily integrable form approximating the nuclear potential through zero-range contact interactions. The spin-orbit terms of the interaction have been neglected, since we are
studying pure neutron matter, a spin-saturated system.

Many Skyrme functionals have been produced and used across the literature. In this work we focus on the commonly used SLy4 and SkM parametrizations, as well as KDE0v1, NRAPR, and SKRA which have been shown to respect a set of constraints coming from neutron matter~\cite{Dutra:2012,Brown:2014}. 

The many-body wave function for $N$ neutrons is approximated by the Slater determinant of the single neutron wave functions $\phi_i(x_j) : i=1,\dots,N$:
\begin{align}
	\phi(x_1, \dots , x_N) = \tfrac{1}{\sqrt{N!}}\text{det}|\phi_i(x_j)| \label{eq:slaterdet}
\end{align}
where $x_j$ denotes the tuple $(\bvec{r}, \sigma)$ of spatial coordinates and spin for a given state. In general, the total energy of the system is given by the expectation value:
\begin{align}
	E &= \bra{\phi}\hat{H}\ket{\phi} \nonumber\\
	&= \int \mathcal{H}(\bvec{r})d^3r \label{eq:energyexpectation}
\end{align}
The energy-density functional $\mathcal{H}(\bvec{r})$ for neutron matter, neglecting spin density terms, is given by the expression: 
\begin{align}
	\mathcal{H} = \mathcal{K} + \mathcal{H}_0 + \mathcal{H}_3 + \mathcal{H}_{\text{eff}} + \mathcal{H}_{\text{fin}}
\end{align}
Each of these terms represents a different component of the nuclear Hamiltonian:
\begin{align}
	&\mathcal{K} = \frac{\hbar^2}{2m}\tau(\bvec{r}), \nonumber\\
	&\mathcal{H}_0 =\big(C_0^{\rho,0}+C_1^{\rho,0}\big)\rho^2(\bvec{r})=\frac{1}{4}t_0(1-x_0)\rho^2(\bvec{r}), \nonumber\\
	&\mathcal{H}_3 = \big(C_0^{\rho,\alpha}+C_1^{\rho,\alpha}\big)\rho^{2+\alpha}(\bvec{r})=\frac{1}{24}t_3(1-x_3)\rho^{2+\alpha}(\bvec{r}),\nonumber\\
	&\mathcal{H}_{\text{eff}} = \big(C_0^{\tau}+C_1^{\tau}\big)\rho(\bvec{r})\tau(\bvec{r})\nonumber\\
	&= \frac{1}{8}\big[t_1(1-x_1)+3t_2(1+x_2)\big]\rho(\bvec{r})\tau(\bvec{r}),\, {\rm and} \nonumber\\
	&\mathcal{H}_{\text{fin}} = -\big(C_0^{\Delta\rho}+C_1^{\Delta\rho}\big)\big(\nabla\rho(\bvec{r})\big)^2\nonumber\\
	&=\frac{3}{32}\big[t_1(1-x_1)-t_2(1+x_2)\big]\big(\nabla\rho(\bvec{r})\big)^2 \nonumber\\ 
\end{align}
are the kinetic energy, zero-range, three-body, effective mass, and finite-range interactions respectively. The $C_i$ coefficients are the Skyrme parameters in the isospin-representation basis and are defined in terms of the original Skyrme parameters in Appendix A.

The definitions of the neutron density $\rho(\bvec{r})$ and kinetic density $\tau(\bvec{r})$ follow from the definition of the Slater determinant wavefunction and are expressed in terms of the single particle wavefunctions $\phi_i(\bvec{r},\sigma)\equiv \langle\bvec{r},\sigma|\phi_i\rangle$:
\begin{align}
	\rho(\bvec{r}) = \sum_{i,\sigma}|\phi_i(\bvec{r},\sigma)|^2\nonumber\\
	\tau(\bvec{r}) = \sum_{i,\sigma}|\bvec{\nabla}\phi_i(\bvec{r},\sigma)|^2
	\label{eq:densities}
\end{align}
where the $i$ index runs from $1$ to $N$ and $\sigma$ sums over spin-up and spin-down components. 

Without any perturbation, the ground-state orbitals for this energy-density functional are given by plane-waves. That is, the $\phi_i$ are the same as that of the non-interacting gas. Thus we have:
\begin{align}
\rho(\bvec{r}) &= \sum_{i,\sigma}|\phi_i(\bvec{r},\sigma)|^2=\rho_0
\end{align}
and
\begin{align}
\tau(\bvec{r}) &= \sum_{i,\sigma}|\bvec{\nabla}\phi_i(\bvec{r},\sigma)|^2=2\sum_{|\bvec{k}_i|<=k_F}\frac{|\bvec{k}_i|^2}{V}\nonumber\\
&=\rho_0 \bigg(2 \sum_{|\bvec{k}_i|<=k_F}\frac{|\bvec{k}_i|^2}{N}\bigg)=\rho_0\frac{\bar{E}_{FG}}{\big(\frac{\hbar^2}{2m}\big)}\nonumber\\
&=\frac{3}{5}(3\pi^2)^{2/3}\rho_0^{5/3}
\end{align}
where the kinetic density is given in the TL. For a finite number of particles one would sum over the occupied $\bvec{k}$ states instead. Since the densities are constants across space the energy is given by
\begin{align}
E=\mathcal{H}(\rho_0)V
\end{align}
and the energy per particle is
\begin{align}
\bar{E}(\rho_0)=\frac{\mathcal{H}(\rho_0)}{\rho_0}
\end{align}
in the TL.

\begin{figure}[t]
\begin{center}
\includegraphics[width=1.0\columnwidth,clip=]{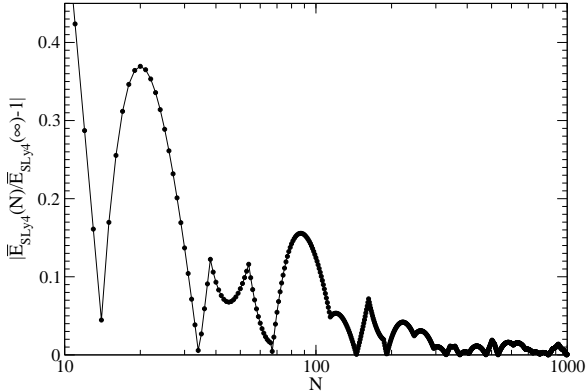}
\caption[Skyrme SLy4 FS effects]{Relative error of N particle energy compared to TL energies shows magnitude of finite size effects. SLy4 parameterization at $\rho_0 = 0.1\,\rm{fm^{-3}}$. \label{fig:SLy4 TL}}
\end{center}
\end{figure}

One may perform a FS study similarly to the non-interacting gas. A qualitative analysis shows a minimum in FS effects for SLy4 in the energy per particle around $66$ particles at $\rho_0=0.1\,{\rm fm}^{-3}$. This can be seen in Fig.~\ref{fig:SLy4 TL} which has the same trend as Fig.~\ref{fig:FFG-TL}. Mid-shell, the Skyrme results exhibit much
larger finite-size effects than the noninteracting Fermi gas; it is precisely such
interaction-based effects that we will be interested in capturing in what follows
(though we will now also turn on an external potential, thereby complicating the
calculation).

With the choice of our external potential, the energy density functional becomes:
\begin{align}
	\mathcal{H}_v(\bvec{r}) = \mathcal{H}(\bvec{r}) + 2v_q\cos(qz)\rho(\bvec{r})
	\label{eq:fullEDF}
\end{align}
Solving for the ground-state energy is more difficult for the inhomogeneous problem and is done via a self-consistent mean-field approach described in the next section.

\subsection{Hartree-Fock equations}
\label{sec:hartreefock}
Hartree-Fock theory is used to minimize the energy within a certain space of wave-functions. This involves the Slater determinant of single-particle orbitals as outlined in Eq.~(\ref{eq:slaterdet}) that satisfies antisymmetry properties required for fermionic systems.
The total energy is minimized with respect to variations in the single-particle wavefunctions through the method of Lagrange multipliers:
\begin{align}
	&\frac{\delta}{\delta\phi_i^*(\bvec{r},\sigma)}\Big(E-\sum_je_j\int d\bvec{r}\sum_{\sigma}|\phi_j(\bvec{r},\sigma)|^2 \Big) \nonumber\\
	&=\frac{\partial\mathcal{H}_v}{\partial\phi_i^*(\bvec{r},\sigma)}-\nabla\cdot\frac{\partial\mathcal{H}_v}{\partial\nabla\phi_i^*(\bvec{r},\sigma)}-e_i\phi_i(\bvec{r},\sigma)
	=0
	\label{eq:hartreefockorigin}
\end{align}
which yields the single-particle Hartree-Fock equations: 
\begin{align}
	- \nabla\cdot\Bigg(\frac{\hbar^2}{2m^*(\bvec{r})}\nabla\phi_i(\bvec{r},\sigma)\Bigg)
	+ U(\bvec{r})\phi_i(\bvec{r},\sigma)\nonumber\\
	+ v(z)\phi_i(\bvec{r},\sigma) = e_i\phi_i(\bvec{r},\sigma)
	\label{eq:hartreefockdef}
\end{align}
Where we define the effective mass term:
\begin{align}
	\frac{\hbar^2}{2m^*(\bvec{r})} = \frac{\hbar^2}{2m}+\frac{1}{8}\big[t_1(1-x_1)+3t_2(1+x_2)\big]\rho(\bvec{r}) 
	\label{eq:effectivemass}
\end{align}
and the interaction potential term:
\begin{align}
   	U(\bvec{r}) &= \frac{1}{2}t_0(1-x_0)\rho(\bvec{r})
	+ \frac{2+\alpha}{24}t_3(1-x_3)\rho^{1+\alpha}(\bvec{r}) \nonumber\\
	&+ \frac{1}{8}\big[t_1(1-x_1)+3t_2(1+x_2)\big]\tau(\bvec{r}) \nonumber\\
	&- \frac{3}{16}\big[t_1(1-x_1)-t_2(1+x_2)\big]\nabla^2\rho(\bvec{r})
	\label{eq:interatctionpotential}
\end{align}
 In addition, it is important to note that due to a lack of spin polarization, Eq.~(\ref{eq:hartreefockdef}) is the same differential equation (DE) for all orbitals regardless of spin. This allows us to choose a solution of orbitals half of which are purely spin up, and the other half which share the same spatial orbital functions as the spin-up states, but are spin down instead. This is the same orbital structure we saw in the noninteracting Fermi gas where we had the same plane-waves occupied twice each due to spin-up and spin-down. We write this as:
\begin{align}
\phi_i(\bvec{r},\downarrow)=0,\, 0 \leq i\leq N/2;\nonumber\\
\phi_i(\bvec{r},\uparrow)=0,\, N/2< i\leq N;\nonumber\\
\phi_i(\bvec{r},\downarrow)=\phi_{i-N/2}(\bvec{r},\uparrow),\,N/2< i\leq N,
\end{align}
where the zero spin components trivially satisfy the HF equation and the double degeneracy of spatial orbitals is allowed because the DE is the same for both spin components so it is valid to use the same spatial solution twice for different spins, as per the Pauli exclusion Principle.

We can thus rewrite the density and Hartree-Fock equations in terms of the $N/2$ unique spatial orbitals for which we use a new notation $\phi_i(\bvec{r})$ and introduce a factor of $2$ for spin degeneracy:
\begin{align}
&\rho(\bvec{r}) = 2\sum_{i=1}^{N/2}|\phi_i(\bvec{r})|^2\nonumber\\
&\tau(\bvec{r}) = 2\sum_{i=1}^{N/2}|\bvec{\nabla}\phi_i(\bvec{r})|^2\nonumber\\
&- \nabla\cdot\Bigg(\frac{\hbar^2}{2m^*(\bvec{r})}\nabla\phi_i(\bvec{r})\Bigg)
+ U(\bvec{r})\phi_i(\bvec{r})
+ v(z)\phi_i(\bvec{r})\nonumber\\
& = e_i\phi_i(\bvec{r})
\label{eq:densities2}
\end{align}

In infinite neutron matter, in the absence of an external potential (i.e. $v_q=0$), the density and kinetic energy density must be homogeneous and isotropic throughout due to the symmetry of the system. Plane-waves for the $\phi_i$ satisfy the homogeneous Hartree-Fock equations, thus validating the statements made in the previous section. However, with the addition of the external potential along the z-axis this symmetry is broken and the number density and kinetic density now depend on the z-coordinate.
\begin{align}
	\rho(\bvec{r}) = \rho(z) \quad,\quad \tau(\bvec{r}) = \tau(z) 
\end{align}
Noticing that the effective-mass term Eq.~(\ref{eq:effectivemass}) and the interaction potential term Eq.~(\ref{eq:interatctionpotential}) only depend on $\bvec{r}$ through the density and kinetic density functions, we will again observe that these terms are also dependent only on the z-coordinate.
\begin{align}
	\frac{\hbar^2}{2m^*(\bvec{r})} = \frac{\hbar^2}{2m^*(z)} \quad,\quad U(\bvec{r}) = U(z)
\end{align}
Expanding Eq.~(\ref{eq:densities2}) and employing the separation of variables technique by decomposing the single particle wavefunction into the form $\phi_i(\bvec{r}) = X_i(x)Y_i(y)Z_i(z)$ the following decoupled second-order ordinary differential equation governing the single-particle wavefunctions is recovered:
\begin{align}
	&\frac{\frac{d^2}{d x^2}X_i(x)}{X_i(x)} 
	+ \frac{\frac{d^2}{d y^2}Y_i(y)}{Y_i(y)} 
	+ \frac{\frac{d^2}{d z^2}Z_i(z)}{Z_i(z)} \nonumber\\
	&+ \Bigg(\frac{\frac{d}{d z}\frac{\hbar^2}{2m^*(z)}}{\frac{\hbar^2}{2m^*(z)}} \Bigg)\frac{\frac{d}{d z}Z_i(z)}{Z_i(z)} 
	-\frac{U(z) + v(z) - e_i}{\frac{\hbar^2}{2m^*(z)}} = 0 
	\label{eq:separatedpde}
\end{align}
In the same manner as the noninteracting-gas treatment, we work in a finite periodic box of volume $L^3$. This imposes the standard periodic boundary conditions:
\begin{align}
	&X_i(x) = X_i(x+L) \nonumber\\&Y_i(y) = Y_i(y+L) \nonumber\\&Z_i(z) = Z_i(z+L)
\end{align}
The solutions in $x$ and $y$ are still plane-waves:
\begin{align}
	X_i(x) = \frac{1}{\sqrt{L}} \exp\big[ik_{i,x}x\big] \;\;,\;\; Y_i(y) = \frac{1}{\sqrt{L}} \exp\big[ik_{i,y}y\big] 
\end{align}
with corresponding wavenumbers:
\begin{align}
	k_{i,x} = \frac{2\pi n_{i,x}}{L} \quad , \quad k_{i,y} = \frac{2\pi n_{i,y}}{L} \quad : \quad n_{i,x}, n_{i,y} \in\mathbb{Z}
\end{align}
The ordinary DE governing the z-axis component wavefunction is then given by
\begin{align}
	&-\frac{\hbar^2}{2m^*(z)}\frac{d^2}{d z^2}Z_i(z)
	- \Bigg[\frac{d}{d z}\frac{\hbar^2}{2m^*(z)} \Bigg]\frac{d}{d z}Z_i(z) \nonumber\\
	&+\Bigg[U(z) + v(z) + \frac{\hbar^2}{2m^*(z)}\frac{4\pi^2}{L^2}(n^2_{i,x} + n^2_{i,y})\Bigg]Z_i(z)  
	= e_iZ_i(z)  
	\label{eq:zaxiseq}
\end{align}
Note that since Eq.~(\ref{eq:zaxiseq}) depends on the sum $n_{x}^2+n_{y}^2$, we will have to solve several DE's for various values of this sum in order to find all the orbitals of the ground-state. We define $S(x,y)\equiv n_x^2+n_y^2$. We denote the z-orbital solutions to the DE coresponding to $S$ by $\phi_i^{S}(z)$ and their eigenvalues by $e_i^S$. The set of DE's that need to be solved are:
\begin{align}
	&-\frac{\hbar^2}{2m^*(z)}\frac{d^2}{d z^2}\phi_i^S(z)
	- \Bigg[\frac{d}{d z}\frac{\hbar^2}{2m^*(z)} \Bigg]\frac{d}{d z}\phi_i^S(z) \nonumber\\
	&+\Bigg[U(z) + v(z) + \frac{\hbar^2}{2m^*(z)}\frac{4\pi^2}{L^2}S\Bigg]\phi_i^S(z)  
	= e_i^S \phi_i^S(z)  
	\label{eq:zaxiseq2}
\end{align}
where the maximal relevant value of $S$ depends on the number of particles $N$.

\subsection{Finite-difference method}
\label{sec:finitediff}
We now consider the numerical methods to be used in computing the single particle wavefunctions and energies. In our case, this reduces to solving a second-order linear ordinary differential equation with periodic boundary conditions, given by Eq.~(\ref{eq:zaxiseq2}). We relabel the coefficients:
\begin{align}
&A(z)=-\frac{\hbar^2}{2m^*(z)},\nonumber\\
&B(z)=- \Bigg[\frac{d}{d z}\frac{\hbar^2}{2m^*(z)} \Bigg],\nonumber\\
&C^S(z)=\Bigg[U(z) + v(z) + \frac{\hbar^2}{2m^*(z)}\frac{4\pi^2}{L^2}S\Bigg],\nonumber\\
&A(z)\frac{d^2}{d z^2}\phi^{S}_i(z) + B(z)\frac{d}{d z}\phi_{i}^S(z) + C^S(z)\phi_{i}^S(z) = e_i^S\phi_i^S(z)
	\label{eq:generalode}
\end{align}
with periodic boundary conditions
\begin{align}
&\phi_i^S(z) = \phi_i^S(z+L);\nonumber\\
&A(z)=A(z+L),\,B(z)=B(z+L),\,C^S(z)=C^S(z+L)
\end{align}
We introduce a finite difference scheme \cite{Hairer:2008} by discretizing the space. For a periodic box of length $L$, the single-particle orbital $\phi_i^S(z)$ is discretized and written in vector form:
\begin{align}
	\bvecg{\phi}_{i}^S =
	\begin{pmatrix}
		[\phi_{i}^S]^0 \\
		\\
		\vdots\\
		\\
		[\phi_{i}^{S}]^{M-1} \\
	\end{pmatrix}
	\quad:\quad [\phi_{i}^{S}]^j \equiv \phi_{i}^S(j\Delta z) \quad,\quad \Delta z = \frac{L}{M}
\end{align}
where $M$ is the number of equally spaced points in the box where the orbitals are evaluated at. This discretizes the entire problem including the density quantities which are functions of the orbitals and their derivatives so that:
\begin{align}
&\rho^j\equiv \rho(j\Delta z),\, \tau^j\equiv \tau(j\Delta z),\,\nonumber\\
&A^j\equiv A(j\Delta z),\, B^j\equiv B(j\Delta z),\, [C^S]^j\equiv C^S(j\Delta z)
\end{align}
With the basic discretization scheme in place, a 5-point stencil is used in approximating the first and second derivatives of the orbitals. They are given by:
\begin{align}
    \frac{d}{d z}\phi_i^S(z)&=\frac{1}{12\Delta z}\Big[-\phi_i^S(z+2\Delta z)+8\phi_i^S(z+\Delta z)\nonumber\\
    &-8\phi_i^S(z-\Delta z)+\phi_i^S(z-2\Delta z)\Big]+\mathcal{O}(\Delta z^4)\nonumber\\
	\frac{d^2}{d z^2}\phi_{i}^S(z) 
	&= \frac{1}{12\Delta z^2}\Big[-\phi_i^S(z+2\Delta z)+16\phi_i^S(z+\Delta z)  \nonumber\\
	&-30\phi_i^S(z)+16\phi_i^S(z-\Delta z)-\phi_i^S(z-2\Delta z)\Big]  \nonumber\\
	&+ \mathcal{O}(\Delta z^4)
	\label{eq:phi2der}
\end{align}
Using these approximations, one can replace the derivatives in the DE with the orbitals evaluated at the 5 locations of the stencil. Thus the discretized version of the DE is:
\begin{align}
	&\Bigg(-\frac{A^j}{12\Delta z^2}-\frac{B^j}{12\Delta z}\Bigg)[\phi_i^S]^{\rm{mod}(j+2,M)}\nonumber\\
	&+\Bigg(\frac{16A^j}{12\Delta z^2}+\frac{8B^j}{12\Delta z}\Bigg)[\phi_i^S]^{{\rm mod}(j+1,M)}\nonumber\\
	&+\Bigg([C^S]^j-\frac{30A^j}{12\Delta z^2}\Bigg)[\phi_i^S]^{j} \nonumber\\
	&+\Bigg(\frac{16A^j}{12\Delta z^2}-\frac{8B^j}{12\Delta z}\Bigg)[\phi_i^S]^{\rm{mod}(j-1,M)}\nonumber\\
	&+\Bigg(-\frac{A^j}{12\Delta z^2}+\frac{B^j}{12\Delta z}\Bigg)[\phi_i^S]^{\rm{mod}(j-2,M)} 
	= e_i^S[\phi_i^S]^j
	\label{eq:matrixdefold}
\end{align}
for $0\leq j \leq M-1$. The function $\rm{mod}(K,M)$ returns the remainder of K divided by M in the interval $[0,M-1]$ and is used to enforce the boundary conditions. Equation~(\ref{eq:matrixdefold}) recasts the DE into an matrix eigenvalue problem:
\begin{align}
	F^S\bvecg{\phi}_i^S = e_i^S\bvecg{\phi}_{i}^S
\end{align}
$F^S$ is an $M\times M$ matrix with zon-zero entries located at $F^S_{j,\rm{mod}(K,M)}$ where $K=j-2,\,j-1,\,j,\,j+1,\,\rm{or}\, j+2$ and $0\leq j \leq M-1$, and the matrix element is the corresponding coefficient in front of the $[\phi_i^S]^{\rm{mod}(K,M)}$ term in Eq.~(\ref{eq:matrixdefold}). The solutions to the eigenvector problem are a set of $M$ eigenvectors $\bvecg{\phi}_{i}^S$, $1 \leq i \leq M$ with corresponding eigenvalues $e_i^S$.

Note that while the eigenvectors correspond to the $z$ components of the orbitals, the eigenvalues are the Lagrange multipliers for the entire single-particle state. For no interactions, the $e_i$'s are the single-particle orbital energies from the time-independent Schr\"odinger's equation. With Skyrme interactions, they no longer correspond to the same quantity (i.e. one cannot sum up the $e_i$'s of occupied states to compute the energy of the system). However, the states are still occupied in order of increasing eigenvalue cardinal number.

The goal is to find the $N/2$ smallest eigenvalues and their eigenvectors across all $(n_x,\,n_y)$ pairs. These are the occupied states which we labelled as $Z_i(z),e_i$ for $1 \leq i \leq N/2$. Generally, the larger the value of $S$, the larger the eigenvalues of $F^S$. That is, the quantity $\min(e_i^S)=e_1^S$ increases as $S(x,y)$ increases. Thus, if $e_1^{S'}$ of $F^{S'}$ is greater than the $N/2$ smallest states found thus far, then this implies that $\max(S)<S'$ where $\max(S)$ is the largest $S$ value containing occupied states.

The quality of the eigenvector approximation is best at the minimum eigenvalue and degrades as the cardinal number of the eigenvalue increases. Thus one should work at $M$ large enough to produce enough good-quality orbitals to accommodate all of the particles in the finite system.

\subsection{Iterative solution techniques}
\label{sec:iterative}

We now outline the basic iterative procedure for solving the Hartree-Fock equations. The densities are a function of the orbitals. The effective mass and effective interaction/matrix elements are functions of the densities. This dependence of the densities on the eigenvectors poses a self-consistency problem. Solving the matrix eigenvalue problem changes/updates the orbitals, thus updating the densities. However, this changes the DE/matrix itself. One needs to arrive at a set of orbitals which self-consistently satisfy the eigenvalue problem. This is to say that a fixed-point solution is required, which is defined as one where the densities do not change after solving the eigenvalue problem.

The simplest method, which we use here, is to solve and update the densities iteratively until the density and/or energy converges to some fixed-point value. In this work the iterative procedure is as follows:

\begin{enumerate}
\item The number and kinetic densities are initialized with the values corresponding to the homogeneous problem where $v_q$ is set to $0$. These are the same as the non-interacting densities at the same average number density.
\item The matrices are initialized from the densities and the eigenvalues and eigenvectors are computed. The matrix eigenvalue problems are solved across all $(n_x,\,n_y)$ pairs in order of increasing $S$, starting with $S=0$. As the eigenvalue problems are solved:
\begin{enumerate}
\item The sets of $\bvecg{\phi}_i^S$ are sorted by eigenvalue cardinal number for each $(n_x,\,n_y)$ pair.
\item The sorted sets from each $(n_x,\,n_y)$ pair are combined and re-sorted.
\end{enumerate}
\item The search for states ends once the $N/2$ smallest eigenvalues and their eigenvectors across all $(n_x,\,n_y)$ pairs have been found.
	\item With the sorted list of single-particle spatial orbitals produced and normalized prior to, or at this point, the density $\rho(\bvec{r})$, kinetic density $\tau(\bvec{r})$ and energy per particle $\bar{E}$ are calculated. Taking into account spin degeneracy:
		\begin{align}
			&\rho(z) 
			= \frac{2}{L^2}\sum_{i=1}^{N/2}\Big|Z_i(z) \Big|^2,\nonumber\\
			&\tau(z) = \frac{2}{L^2}\sum_{i=1}^{N/2}
		\Big[\frac{4\pi^2}{L^2}\Big(n_{i,x}^2 + n_{i,y}^2\Big)Z_i^2(z) +\nonumber\\
		&\Big(\frac{d}{dz}Z_i(z)\Big)^2\Big],\,{\rm and}\nonumber \\
			&\bar{E} = \frac{1}{N}\int_V d\bvec{r}\mathcal{H}_v(\bvec{r})
		\end{align}

	\item Steps 2 to 4 are repeated with the newly calculated densities to produce the matrices. This is done until the energy per particle has converged to within a specified tolerance or has reached a minimum, corresponding to the completion of the energy minimization procedure.
\end{enumerate}

\begin{figure}[t]
\begin{center}
\includegraphics[width=1.0\columnwidth,clip=]{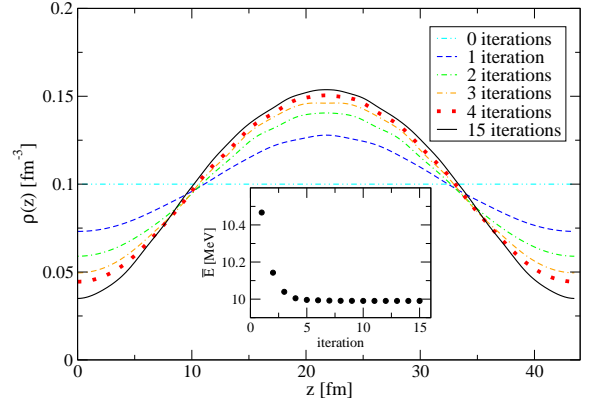}
\caption{SLy4 density of $8250$ particles with external potential parameters: $2v_q = 0.25 E_F$, 1 period in the box and average density $0.10$ fm$^{-3}$.\label{fig:SLy4_n0.1_N8250_1per_0.25EF_density}}
\end{center}
\end{figure}

\begin{figure}[b]
\begin{center}
\includegraphics[width=1.0\columnwidth,clip=]{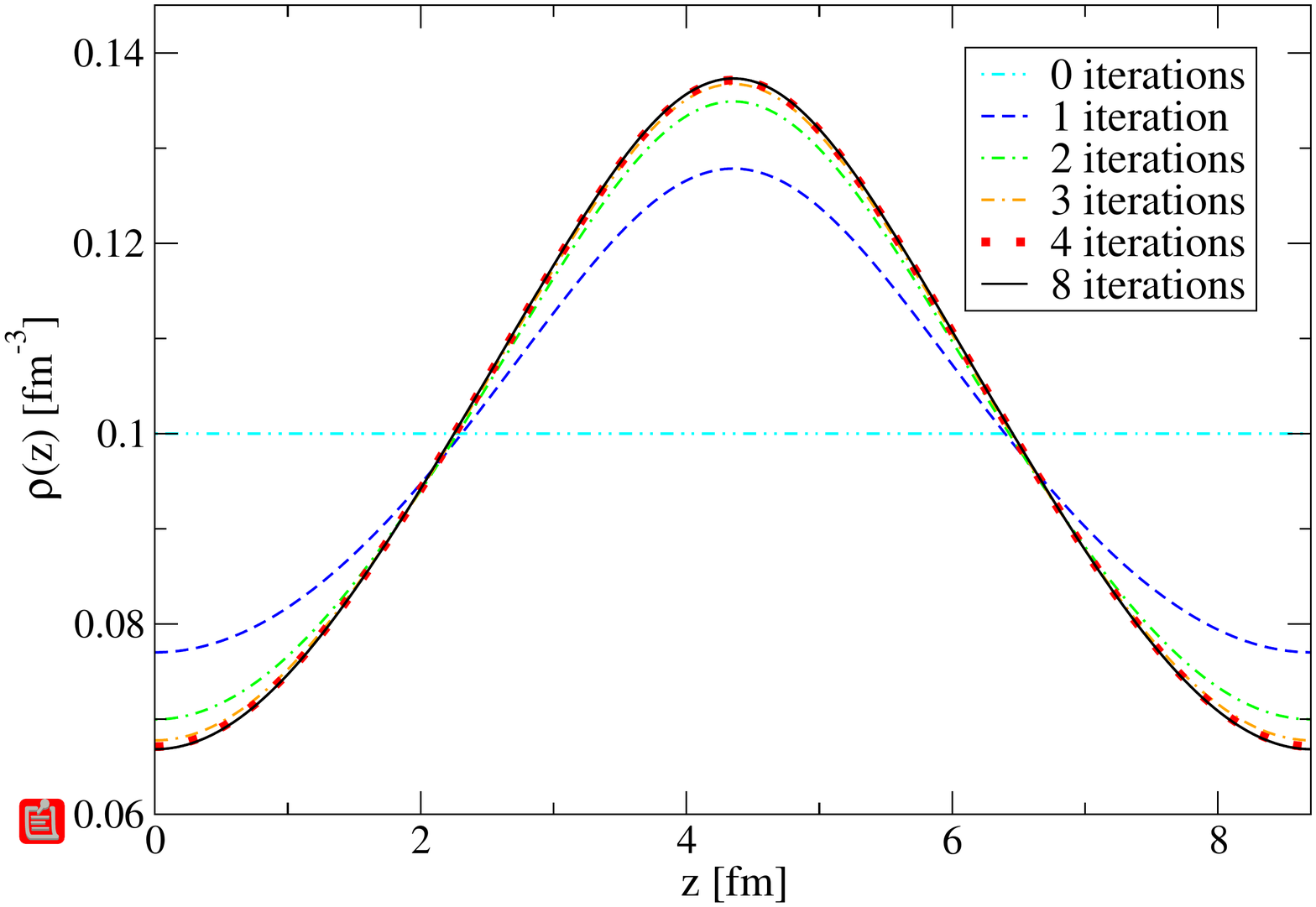}
\caption{SLy4 density of $66$ particles with external potential parameters: $2v_q = 0.25 E_F$, 1 period in the box and average density $0.10$ fm$^{-3}$.\label{fig:SLy4_n0.1_N66_1per_0.25EF_density}}
\end{center}
\end{figure}

\begin{figure}[t]
\begin{center}
\includegraphics[width=1.0\columnwidth,clip=]{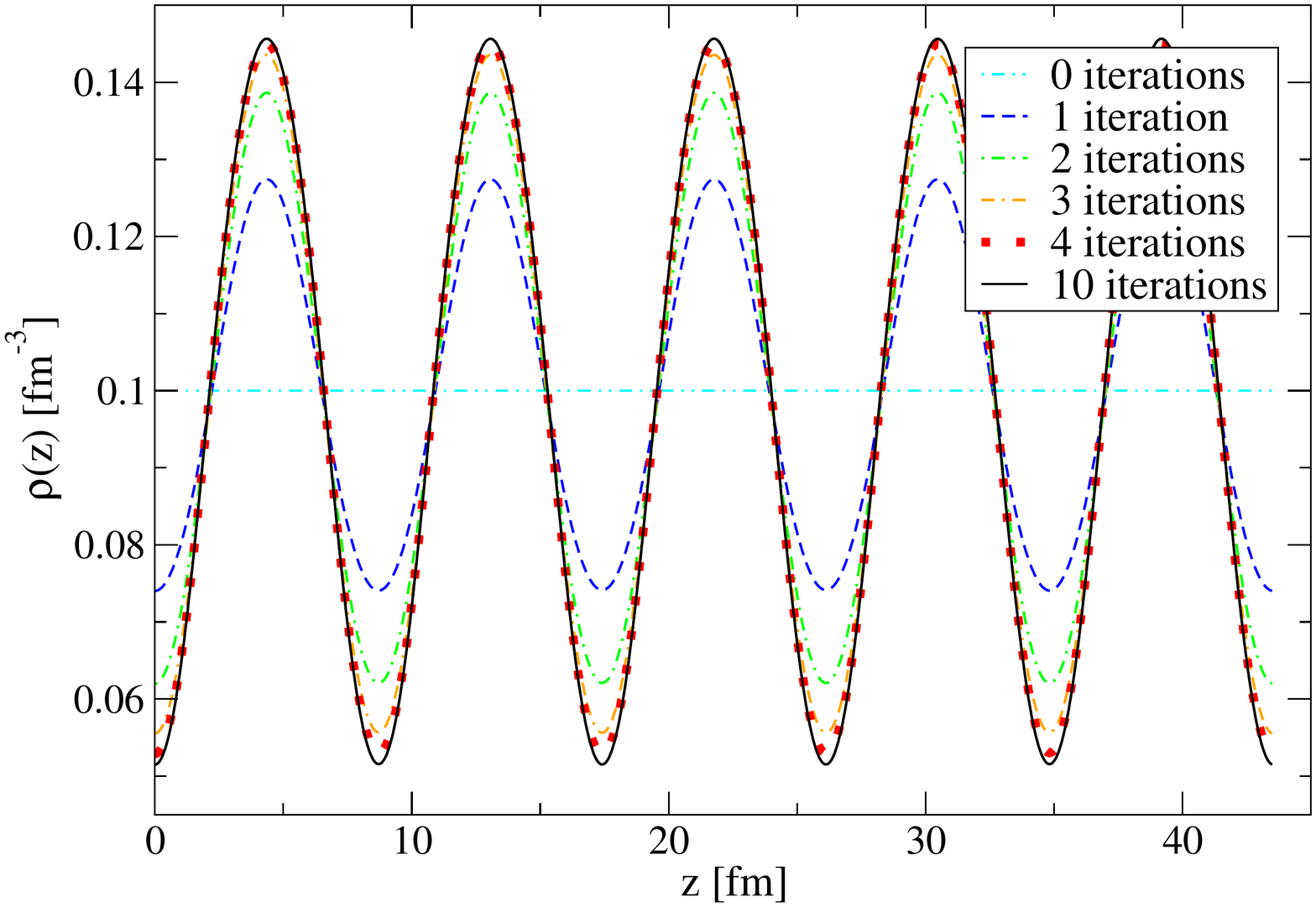}
\caption{SLy4 density of $8250$ particles with external potential parameters: $2v_q = 0.25 E_F$, 5 periods in the box and average density $0.10$ fm$^{-3}$.\label{fig:SLy4_N8250_5per_0.25EF_0.10density}}
\end{center}
\end{figure}

\begin{figure}[b]
\begin{center}
\includegraphics[width=1.0\columnwidth,clip=]{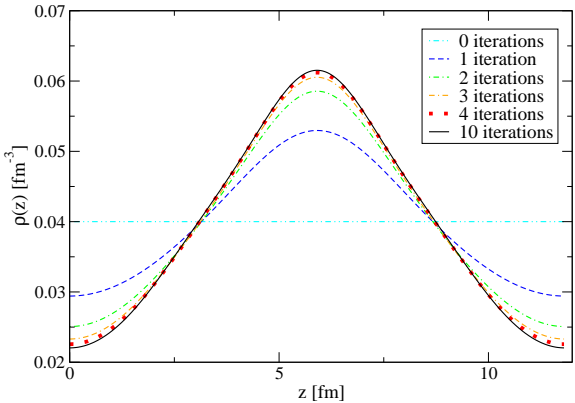}
\caption{SLy4 density of $66$ particles with external potential parameters: $2v_q = 0.25 E_F$, 1 period in the box and average density $0.04$ fm$^{-3}$.\label{fig:SLy4_n0.04_N66_1per_0.25EF_density}}
\end{center}
\end{figure}

The iterative procedure trivially agrees with the analytic solutions in the homogeneous case since  those solutions are self-consistent in the Hartree-Fock equations. The purpose of this algorithm is to solve for the energy in the inhomogeneous $v_q \neq 0$ case. The convergence in the number density for the case of an (unperturbed)
density of $0.10$ fm$^{-3}$ is shown in Fig.~\ref{fig:SLy4_n0.1_N8250_1per_0.25EF_density} for 8250 particles using
the SLy4 parametrization; the inset shows
that the energy settles down to a given value as the iteration count increases.
In Fig.~\ref{fig:SLy4_n0.1_N66_1per_0.25EF_density} we show results at the same
density for the same Skyrme functional, this time for 66 particles. You can 
see that the number of iterations required increases with the particle number.
Note also that the specific form of the (converged) inhomogeneous density 
looks quite different in the two cases: this is to be expected, as the spatial integral
of the number density gives the total number of particles. 
In order to illustrate our ability
to handle different periodicities as well, Fig.~\ref{fig:SLy4_N8250_5per_0.25EF_0.10density} shows results
at the same density for 8250 particles using
the SLy4 parametrization when 5 periods fit inside the box.

\begin{figure}[t]
\begin{center}
\includegraphics[width=1.0\columnwidth,clip=]{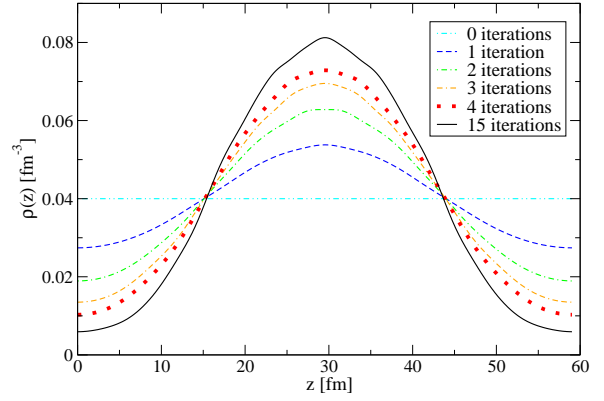}
\caption{SLy4 density of $8250$ particles with external potential parameters: $2v_q = 0.25 E_F$, 1 period in the box and average density $0.04$ fm$^{-3}$.\label{fig:SLy4_n0.04_N8250_1per_0.25EF_density}}
\end{center}
\end{figure}

\begin{figure}[b]
\begin{center}
\includegraphics[width=1.0\columnwidth,clip=]{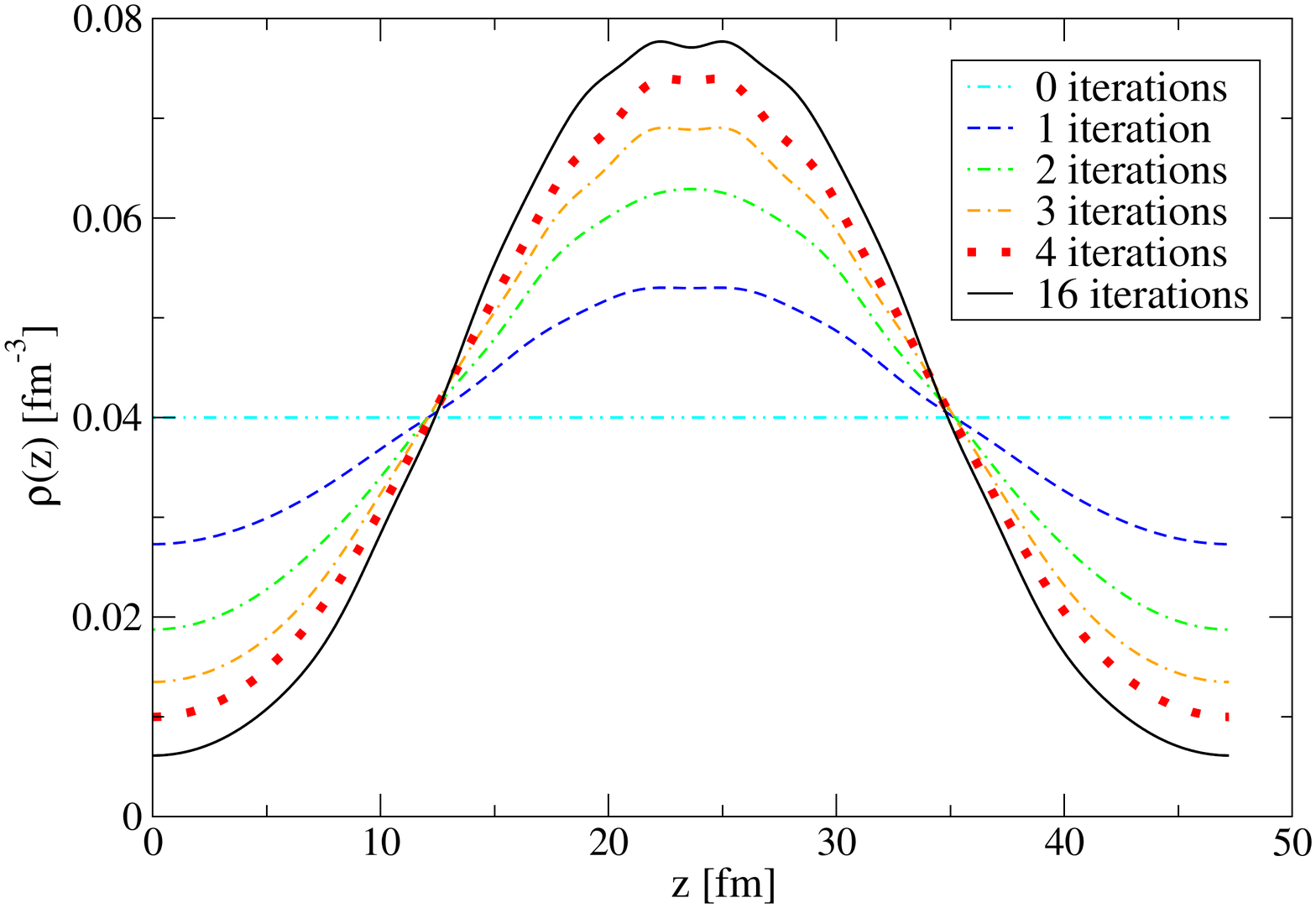}
\caption{SLy4 density of $4224$ particles with external potential parameters: $2v_q = 0.25 E_F$, 1 period in the box and average density $0.04$ fm$^{-3}$.\label{fig:SLy4_n0.04_N4224_1per_0.25EF_density}}
\end{center}
\end{figure}

\begin{figure}[t]
\begin{center}
\includegraphics[width=1.0\columnwidth,clip=]{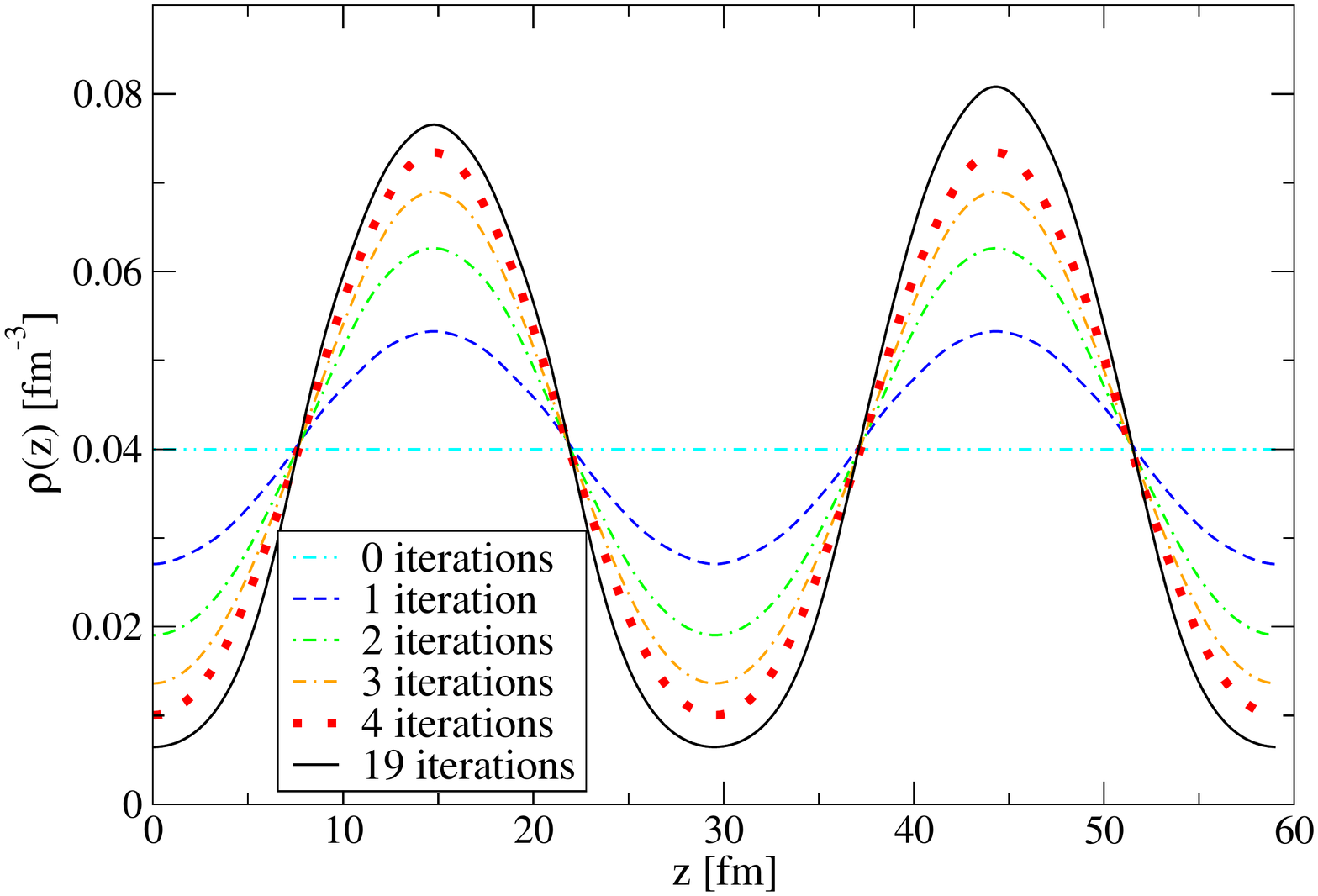}
\caption{SLy4 density of $8250$ particles with external potential parameters: $2v_q = 0.25 E_F$, 2 periods in the box and average density $0.04$ fm$^{-3}$.\label{fig:SLy4_n0.04_N8250_2per_0.25EF_density}}
\end{center}
\end{figure}

\begin{figure}[b]
\begin{center}
\includegraphics[width=1.0\columnwidth,clip=]{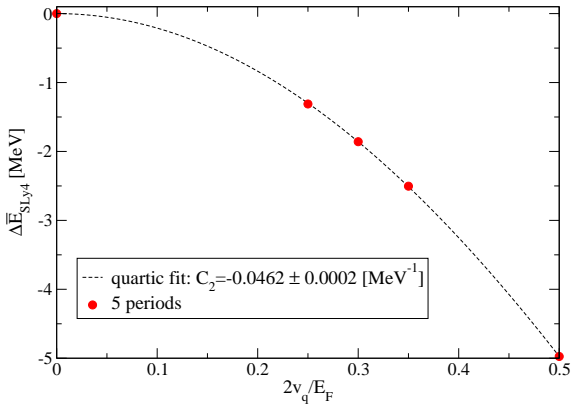}
\caption{SLy4 change in energy per particle versus the potential strength parameter for $8250$ particles at an average density of $0.10$ fm$^{-3}$ with $5$ periods of the potential in the box. The dashed line is the $4^{th}$ degree fit.}\label{fig:SLy4_n0.1_N8250_5per_Evsvqfit}
\end{center}
\end{figure}

The previous three plots were all produced for the same (unperturbed)
density of $0.10$ fm$^{-3}$. We now turn to the case of lower density, 
where we find similar, yet distinct, behavior. Let's start 
from SLy4 results for 1 period in the box involving 66 particles
at an unperturbed density of $0.04$ fm$^{-3}$, shown in Fig.~\ref{fig:SLy4_n0.04_N66_1per_0.25EF_density}: these qualitatively look 
very much like the higher-density 66-particle results of Fig.~\ref{fig:SLy4_n0.1_N66_1per_0.25EF_density}. When we turn to the 
case of 8250 particles at the lower density, see Fig.~\ref{fig:SLy4_n0.04_N8250_1per_0.25EF_density}, we discover as before
that convergence took longer to reach. Intriguingly, the converged
density now exhibits a new feature, a ``dimple'' near the center. 
In an attempt to understand whether this is a numerical artefact,
we went ahead and carried out a corresponding low-density calculation
for 4224 particles, see Fig.~\ref{fig:SLy4_n0.04_N4224_1per_0.25EF_density}: we now find a characteristic ``bubble''-like structure
in the center. As a further probe of what's going on, we also studied the
case of 2 periods fitting in the box, for the same low-density, see Fig.~\ref{fig:SLy4_n0.04_N8250_2per_0.25EF_density}: in this case
(as for that of higher periodicity) the dimple or bubble is no longer present.
That being said, what does appear is an asymmetry between the two peaks, 
an effect which was not present at the higher density, see 
Fig.~\ref{fig:SLy4_N8250_5per_0.25EF_0.10density}. 
This appears to be a numerical instability that arises only for specific
Skyrme parametrizations, only at low density; we have checked and, indeed,
these types of effects are more pronounced at even lower density.

\begin{figure}[t]
\begin{center}
\includegraphics[width=1.0\columnwidth,clip=]{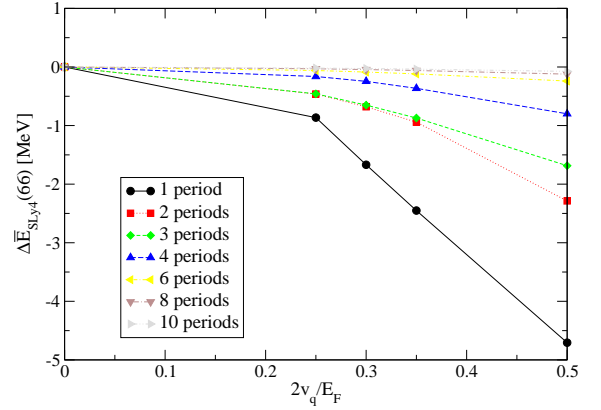}
\caption{SLy4 change in energy per particle versus the potential strength parameter for $66$ particles at an average density of $0.04$ fm$^{-3}$.
Here (and below)
the straight lines serve to guide the eye.
\label{fig:n0.04_SLy4_N66_DeltaEbarvs_vq}}
\end{center}
\end{figure}

\begin{figure}[b]
\begin{center}
\includegraphics[width=1.0\columnwidth,clip=]{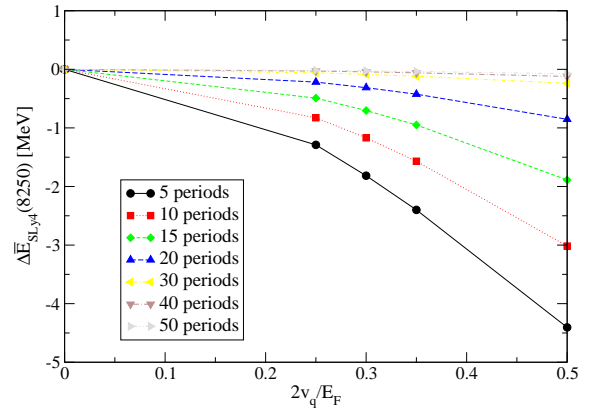}
\caption{SLy4 change in energy per particle versus the potential strength parameter for $8250$ particles at an average density of $0.04$ fm$^{-3}$.\label{fig:n0.04_SLy4_N8250_DeltaEbarvs_vq}}
\end{center}
\end{figure}

It's worth noting that even when the density takes such strange (likely
unphysical) shapes, the energy stays converged. In other words, when setting
up this computational framework, it would have certainly been
possible for us to miss these features, had we merely focused on the 
energy evaluation. Speaking of which,
we have found that the densities and energies may diverge away from their initial settling point if too many iterations are carried out. 
This (distinct) numerical instability is avoided by limiting the number of iterations and specifying a tolerance for convergence that is not too small. It is later shown that response calculations at large particle numbers using such energies agree well with the analytic response in the TL
(the instabilities do not arise for 66 particles). Thus, 
despite the complications discussed here, the results should be valid well within our allowed error tolerance.

\section{Skyrme response}

\begin{figure*}[t]
\begin{center}
\includegraphics[width=\textwidth,height=5cm]{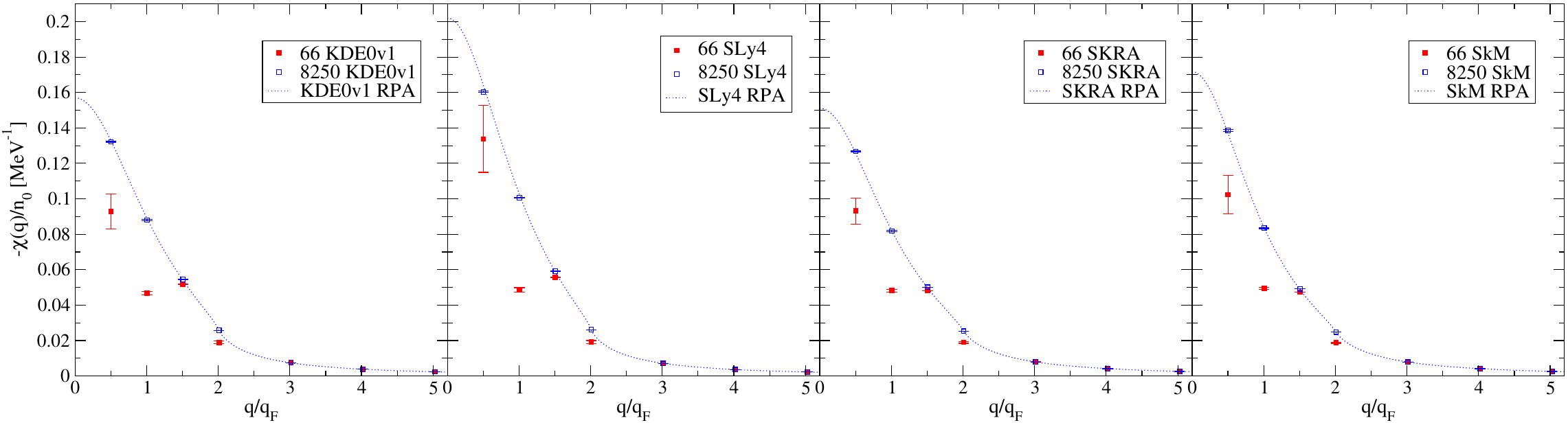}
\caption{Linear response functions of the Skyrme parametrizations KDE0v1, SLy4, SKRA, and SkM (left to right) at a density of $0.04\,{\rm fm}^{-3}$. The dotted lines are the response in the TL. Solid and hollow squares correspond to 66 and 8250 particle responses respectively.}\label{fig:skyrmepanels}
\end{center}
\end{figure*}

The $66$ and $8250$ particle response functions yield important qualitative information about FS effects. This information comes from the change in energy per particle versus $v_q$ data points that are obtained using the methods described in the previous section. We computed energies at the same sets of periodicities and potential strengths as described in the noninteracting-gas response section. For $66$ particles the periodicites are: $1$, $2$, $3$, $4$, $6$, $8$, and $10$ periods in the box. For $8250$ particles, a system with $5$ times the linear dimensions of the $66$ particle system, these exact same periodicities correspond to: $5$, $10$, $15$, $20$, $30$, $40$, and $50$ periods in the larger box respectively. In both cases, these sets of periodicities respectively correspond to $q\approx$ $0.5$, $1.0$, $1.5$, $2.0$, $3.0$, $4.0$, and $5.0$ $q_F$.  In all cases, energies are computed at potential strengths of $2v_q=$ $0.25$, $0.3$, $0.35$, and $0.5$ $E_F$. Thus, the response functions are computed by fitting four $\Delta\bar{E}$ versus $v_q$ points to $\Delta\bar{E}\approx C_2v_q^2+C_4v_q^4$ where $C_2=\chi(q)/\rho_0$.

\begin{figure}[b]
\begin{center}
\includegraphics[width=1.0\columnwidth]{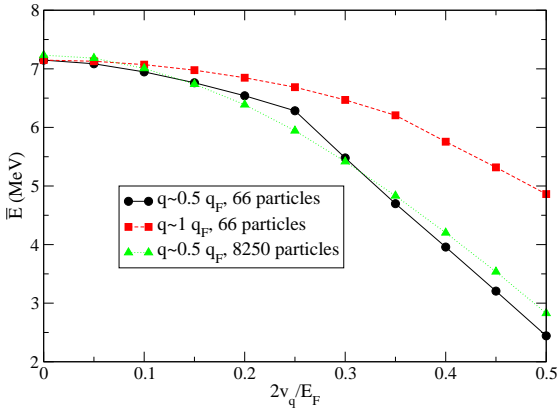}
   \caption{Energy per particle versus external potential strength for SLy4 at a density of $0.04\,{\rm fm}^{-3}$. The solid (circles) and dashed line (squares) correspond to 66 particles with 1 and 2 periods of the potential in the box respectively. The dotted line (triangles) correspond to 8250 particles with the same periodicity as the solid line system.}
\label{fig:slope_66_8250_SLy4_n0.04}
\end{center}
\end{figure}

The magnitude of the response is roughly related to the degree of concavity in the energy versus $v_q$ data sets. An example fit is given for SLy4 at an average density of $0.1\,{\rm fm}^{-3}$ for $8250$ particles with $5$ periods of the potential in the box. The change in energy per particle versus $v_q$ points are given in Fig.~\ref{fig:SLy4_n0.1_N8250_5per_Evsvqfit} alongside a least-squares fit to the quartic form described. The leading coefficient of $-0.0462 \pm 0.0002$ ${\rm MeV}^{-1}$ is an estimate for the linear static-response function at $q \approx 0.5 q_F$.

The degree of curvature in the energy curves is largest at low periodicities and flattens out to $0$ at high periodicities. This can be seen in Figs.~\ref{fig:n0.04_SLy4_N66_DeltaEbarvs_vq} and~\ref{fig:n0.04_SLy4_N8250_DeltaEbarvs_vq} which show energy curves across periodicities for SLy4 at $\rho_0=0.04\,{\rm{fm}}^{-3}$ for both $66$ and $8250$ particles. Even now, when we haven't yet shown any Skyrme-based response 
functions, this behavior is expected: we now that the Lindhard function dies
off as the $q$ is increased, see Fig.~\ref{fig:free0.1resp}.

\begin{figure}[b]
\begin{center}
\includegraphics[width=1.0\columnwidth]{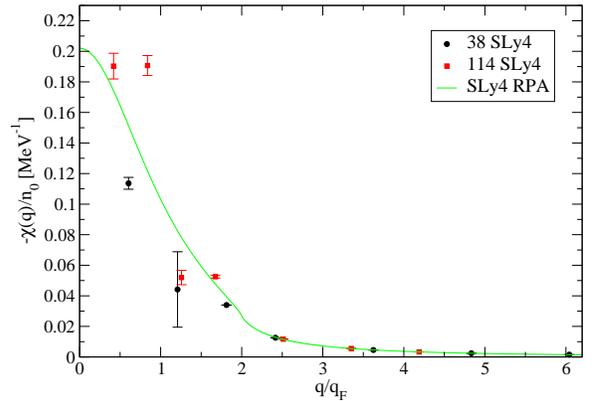}
   \caption{$0.04\,{\rm fm}^{-3}$ SLy4 response functions for different particle numbers. The circles and squares and line correspond to 38 particles, 114 particles, and the TL respectively.}
\label{fig:N38_N114_n0.04response}
\end{center}
\end{figure}

Speaking of which, we are now ready to extract the static-response functions from 
our energy results. 
The response function corresponding to these energy curves is shown on the second panel from the left in Fig.~\ref{fig:skyrmepanels}, which shows the responses extracted from KDE0v1, SLy4, SKRA, and SkM from left to right respectively. The response magnitude is largest at $q \approx 0.5q_F$ and drops down to $0$ to at large $q$. The $66$ and $8250$ particle response s are denoted by solid and hollow squares respectively. The dashed line shows the Skyrme response in the TL, obtained using the random phase approximation (RPA) (see appendix A). The importance of not truncating the fit too early can be seen in the $66$ particle SLy4 response at $2$ ($q \approx 1.0 q_F$) and $3$ ($q \approx 1.5 q_F$) periods. The energy versus $v_q$ curve shows greater curvature at the smaller $q$, but the response magnitude is larger at the bigger $q$ value. If the fit was truncated to the quadratic term the estimates would have the reverse ordering. The general trends are the same across these Skyrme functionals. The $8250$ particle responses show good agreement with the TL response. The FS effects, quantified by the difference between $66$ and TL responses, are largest at the two smallest periodicities and vanish at high $q$.

\begin{figure}[t]
\begin{center}
\includegraphics[width=1.0\columnwidth]{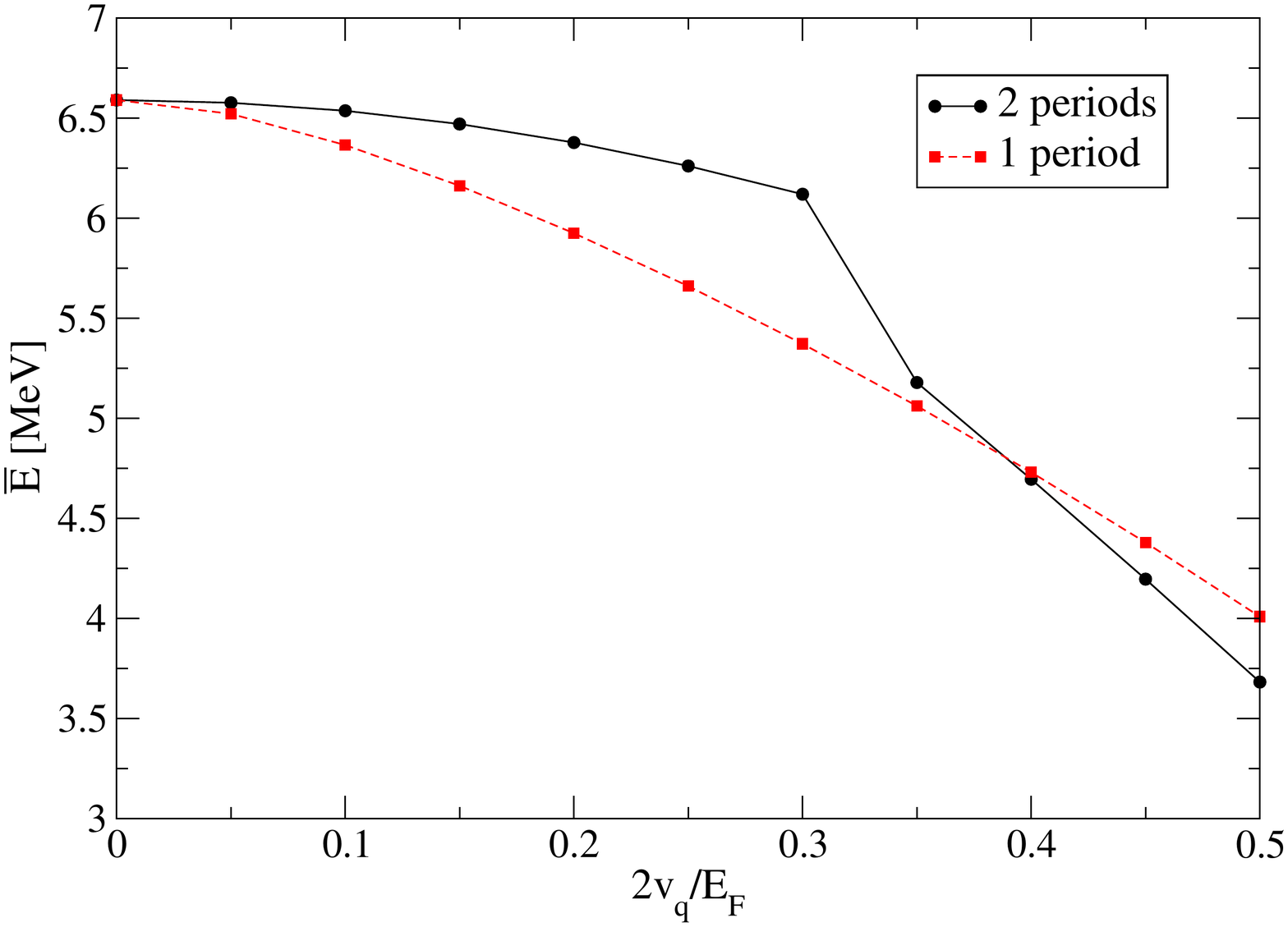}
   \caption{$0.04\,{\rm fm}^{-3}$ energy versus external potential strength for 38 particles for the two smallest periodicities. The circles and squares correspond to one and two periods of the potential traversing the box.}
\label{fig:slope_N38_n0.04response}
\end{center}
\end{figure}

\begin{figure}[b]
\begin{center}
\includegraphics[width=1.0\columnwidth,clip=]{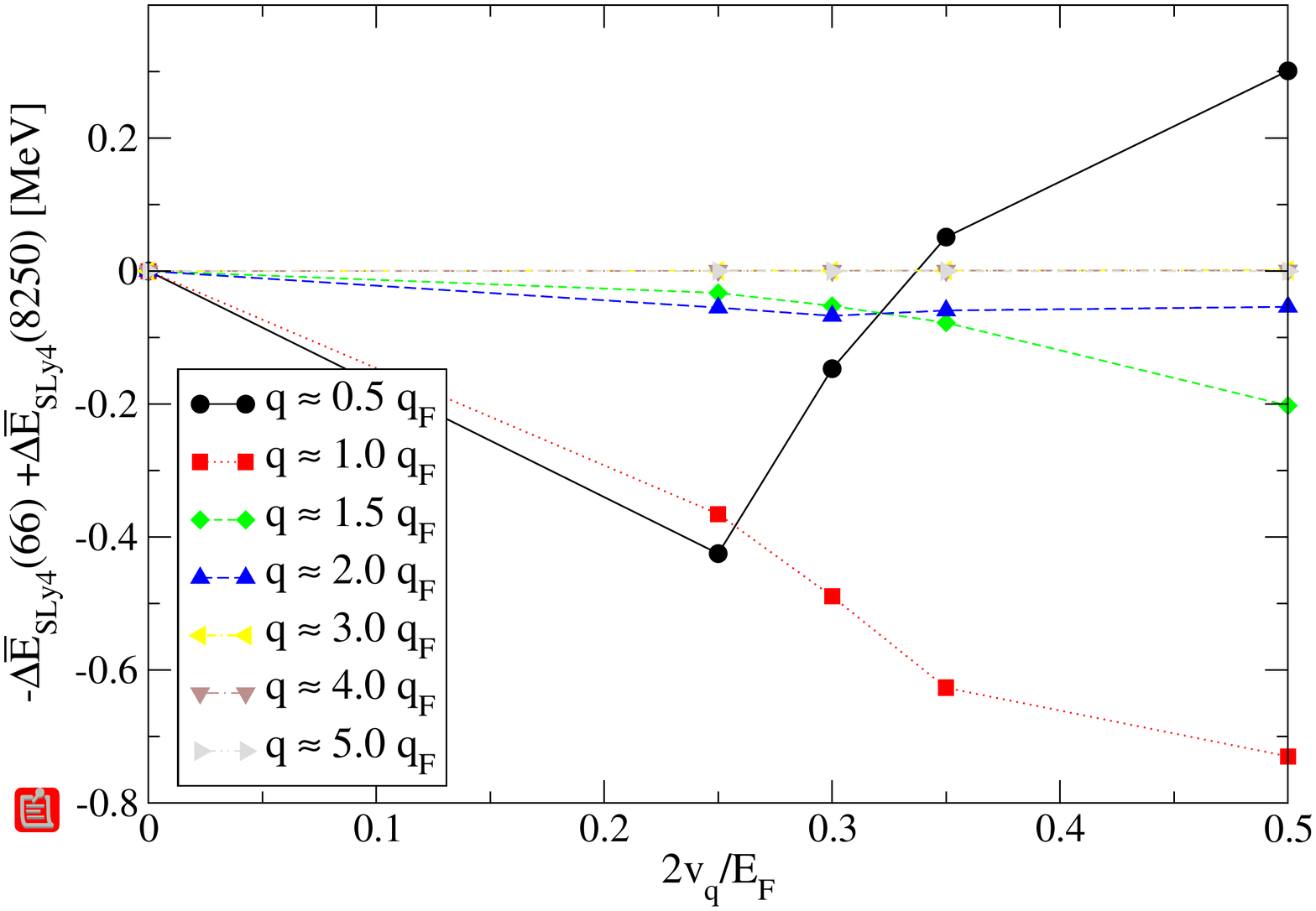}
\caption{SLy4 change in energy per particle contribution to the FS fix versus the potential strength parameter at an average density of $0.04$ fm$^{-3}$.\label{fig:n0.04_SLy4_FSfix_DeltaEbarvs_vq}}
\end{center}
\end{figure}

The large error bars on the lowest-$q$ $66$-particle response points are due to changes in the set of lowest-energy orbitals. This change is induced at certain values of $v_q$ and may reveal itself as a sudden change in slope in the energy versus $v_q$ data. This is shown in Fig.~\ref{fig:slope_66_8250_SLy4_n0.04} where the $66$-particle energy versus $v_q$ data at $1$ period (circles) suddenly changes slope around $2v_q=0.25 E_F$. This results in the large error bar at $q \approx 0.5 q_F$ in the SLy4 response. This is not an issue at $8250$ particles because there are considerably more states and thus a smaller fraction of total states are changed with increasing $v_q$. Indeed the energy versus $v_q$ curve for $8250$ particles at this periodicity (triangles) is smooth. Other $q$ values for $66$ particles have small errors and this is also reflected by smooth energy versus $v_q$ curves. For example, see $q\approx 1 q_F$ (squares) in Fig.~\ref{fig:slope_66_8250_SLy4_n0.04}.

\begin{figure}[t]
\begin{center}
\includegraphics[width=1.0\columnwidth,clip=]{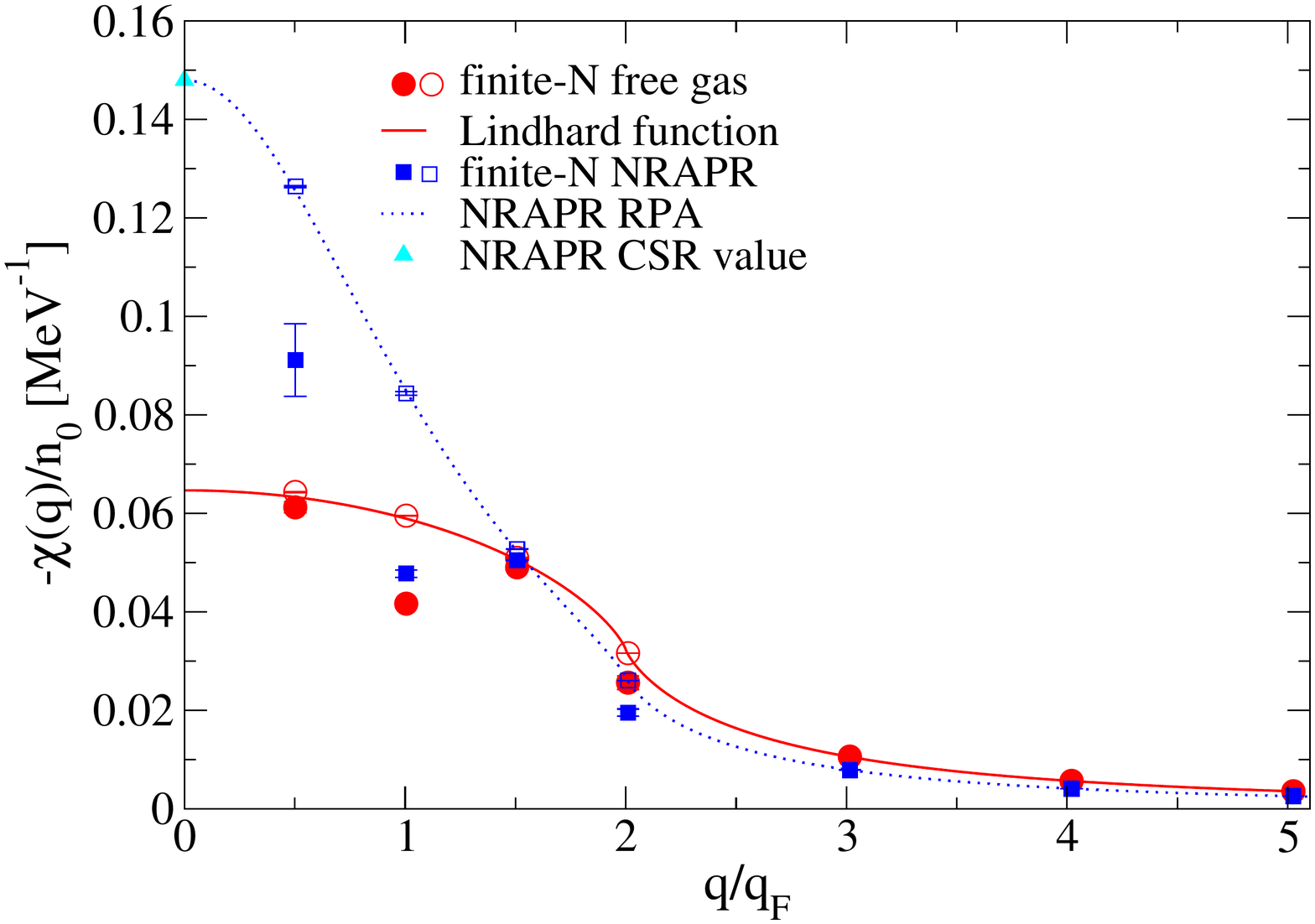}
\caption{Free-gas (circles) and NRAPR (squares) response functions at a density of $0.04\,{\rm fm}^{-3}$. Solid and hollow symbols correspond to 66 and 8250 particles respectively. The solid and dotted lines are the corresponding TL responses of the non-interacting and NRAPR systems.}\label{fig:NRAPR_free_n0.04_resp}
\end{center}
\end{figure}

In addition, the reason why the error is largest at this specific $q$ value is a consequence of the number of particles employed. In an investigation of $38$ particles, the largest error bar for SLy4 occurs at $q\approx 1 q_F$. This is shown in Fig.~\ref{fig:N38_N114_n0.04response} (circles). The energy versus $v_q$ curve in Fig.~\ref{fig:slope_N38_n0.04response} is smooth at 1 period for $38$ particles (squares), resulting in a small error bar. Lastly, it can be seen that FS effects for smaller systems are consistently largest around $q=q_F$. The SLy4 response for $114$ particles (squares) in Fig.~\ref{fig:N38_N114_n0.04response} was included to strengthen this claim for the $3$ separate particle numbers ($38$, $66$, and $114$).

\begin{figure}[b]
\begin{center}
\includegraphics[width=1.0\columnwidth,clip=]{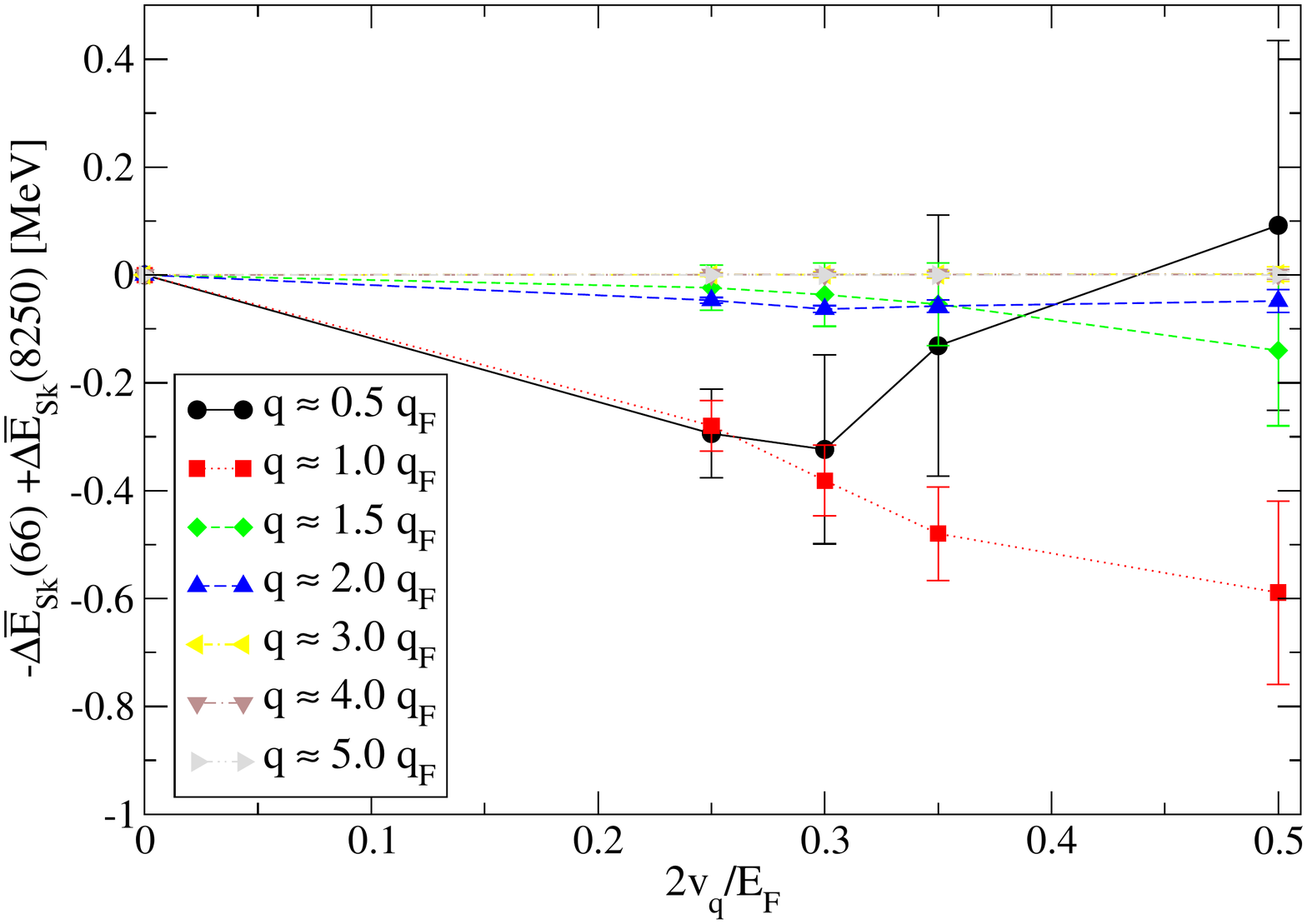}
\caption{Averaged Skyrme change in energy per particle contribution to the FS fix versus the potential strength parameter at an average density of $0.04$ fm$^{-3}$.\label{fig:n0.04_Sk_FSfix_DeltaEbarvs_vq}}
\end{center}
\end{figure}

The prescription for taking FS effects into account applied to Skyrme is:

\begin{align}
\Delta\bar{E}(TL)=\Delta\bar{E}_{\rm{ab~initio}}(66)-\Delta\bar{E}_{Sk}(66)+\Delta\bar{E}_{Sk}(8250)
\label{eq:prescriptionskyrme}
\end{align}
which is to be compared with Eq.~(\ref{eq:prescriptionfree}).
The impact of the term $-\Delta\bar{E}_{Sk}(66)+\Delta\bar{E}_{Sk}(8250)$ on the size of the response is similar to the difference between the $66$ and $8250$ particle Skyrme responses. An example of this fixing term is given in Fig.~\ref{fig:n0.04_SLy4_FSfix_DeltaEbarvs_vq} for SLy4 at a density of $0.04$ $\rm{fm}^{-3}$. The extrapolation to the TL depends on the difference between $66$ and $8250$ particle responses of the system used for FS estimation. Different Skyrme parametrizations will yield slightly different results. A clearer example of this dependence comes from comparing Skyrme with the non-interacting response. Fig.~\ref{fig:NRAPR_free_n0.04_resp} contains finite responses for the non-interacting gas (circles) and the NRAPR Skyrme parametrization (squares). Solid and hollow points correspond to $66$ and $8250$ particles respectively. The TL responses are given by solid and dashed lines respectively. The triangle is the analytic CSR value agreeing with the RPA response. See Appendix A for a discussion on the Skyrme CSR value. The NRAPR response is similar to the other Skyrme responses. One of the most significant differences between the Skyrme and non-interacting responses are the FS effects at the lowest $q$ value. In the noninteracting gas, the $66$ and $8250$ particle responses are practically the same there. Under the FS prescription of Eq.~(\ref{eq:prescriptionfree})
this corresponds to no changes when extrapolating to the TL. However, the Skyrme responses always show sizeable increases in the magnitude of the response moving from $66$ to $8250$ particles at $q\approx 0.5 q_F$ and this is reflected when one
employs a finite-size prescription scheme based on Eq.~(\ref{eq:prescriptionskyrme})
\cite{Buraczynski:2021}. 

To a smaller extent, the results of FS extrapolation are dependent on the specific Skyrme functional employed. In an effort to include all $5$ of the Skyrme functionals employed in this work, we have averaged the energy values of SLy4, SkM, KDE0v1, NRAPR, and SKRA. The FS fix contributions at each periodicity are given in Fig.~\ref{fig:n0.04_Sk_FSfix_DeltaEbarvs_vq}. The differences from the SLy4 FS fix in Fig.~\ref{fig:n0.04_SLy4_FSfix_DeltaEbarvs_vq} are especially apparent at the two lowest periodicities (circles and squares). At each $q$ and $v_q$, the $\Delta\bar{E}_{Sk}(N)$ values for $N=66$ and $N=8250$ from Eq.~(\ref{eq:prescriptionskyrme}) are defined by
\begin{align}
\Delta\bar{E}_{Sk}(N)\equiv\frac{1}{5}\sum_s\Delta\bar{E}_{s}(N)
\end{align}
where the index $s$ runs over SLy4, SkM, KDE0v1, NRAPR, and SKRA.
In addition, we have assigned an uncertainty  to each of these averages:
\begin{align}
{\rm error}\equiv \min\bigg(&\max_s\big(\Delta\bar{E}_{s}(N)\big)-\Delta\bar{E}_{Sk}(N),\nonumber\\
&\Delta\bar{E}_{Sk}(N)-\min_s\big(\Delta\bar{E}_{s}(N)\big)\bigg)
\end{align}
and propagate the errors in  Eq.~(\ref{eq:prescriptionskyrme}) in quadrature.
This is done in an attempt to account for the spread arising from different 
Skyrme parametrizations, i.e., in order not to bias the answer in favor of any given
Skyrme functional. Thus, the average Skyrme fix shown in Fig.~\ref{fig:n0.04_Sk_FSfix_DeltaEbarvs_vq}
is the main result of the formalism developed and applied in this paper;
similar plots could be produced at the other densities we studied.

\section{Summary \& Conclusion}

In conclusion, we examined how neutron matter behaves in the presence of a periodic external field, employing the Hartree-Fock approach for Skyrme energy density functional theory. We constructed a toy model of the complete system by modeling neutron matter first as a non-interacting free Fermi gas and examining the effects the addition of an external potential. This provided insight into multiple areas concerning this study. We were first able to quantify the size of system needed to be able to minimize finite size effects. In contrast to traditional \textit{ab initio} computations that use at most 100-particle systems, we found that the Hartree-Fock methods used allowed for a significantly larger system of 8250 particles to be studied due to the reduction in computational resources needed. We were able to make use of this fact
to greatly reducing finite-size effect contributions by carrying out calculations
for both small and large particle numbers at many external-field strengths and
periodicities.

We were able to compute the linear density-density static-response functions 
for finite particle numbers. These showed very good agreement with analytic results over both a wide range of densities and at high and low periodicities of the external potential. As such, the Skyrme-Hartree-Fock method recommends itself as a robust technique for modelling the infinite neutron matter system, while also providing insights into its structure not achieved through analytic methods. It's worth reiterating that our results, while Skyrme-based, may help to guide \textit{ab initio}
computations for strongly interacting quantum many-body theories, when the latter
attempt to go beyond the simple problem of homogeneous matter.

\begin{acknowledgments}
This work was supported by the Natural Sciences and Engineering Research Council (NSERC) of Canada, the
Canada Foundation for Innovation (CFI), and the Early
Researcher Award (ERA) program of the Ontario Ministry of Research, Innovation and Science. Computational resources were provided by SHARCNET and NERSC.
\end{acknowledgments}

\section{Appendices}

\subsection{Skyrme RPA}

In the random phase approximation applied to Skyrme forces, the linear density-density static-response function for neutron matter is given by \cite{Pastore:2015, Davesne:2009}:
\begin{align}
	\chi_{RPA}(q)&=2\chi_0\Bigg[1 - W_1\chi_0  + W_2\Big(\frac{q^2}{2}\chi_0-2k_F^2\chi_2\Big) \nonumber\\
&+ [W_2]^2k_F^4\Big(-\chi_0\chi_4+\chi_2^2-\frac{q^2}{12\pi^2k_F\big(\frac{\hbar^2}{2m^{*}}\big)}\chi_0\Big)\Bigg]^{-1}
\end{align}
where once again terms coming from the spin-density have been neglected;
a spin-orbit term should in principle be included, but its contribution
is small (vanishing in the limit of zero momentum transfer). The constants showing up in this equation are:
\begin{align}
	\frac{1}{2} W_1
	&= 2\big(C_0^{\rho,0}+C_1^{\rho,0}\big) + (2+\alpha)(1+\alpha)\big[C_0^{\rho,\alpha}+C_1^{\rho,\alpha}\big]\rho^\alpha 
	\nonumber\\&
	-q^2\big[2C_0^{\Delta\rho}+2C_1^{\Delta\rho}+\frac{1}{2}C_0^{\tau}+\frac{1}{2}C_1^{\tau}\big]\nonumber\\
\frac{1}{2} W_2 &= C_0^{\tau} + C_1^{\tau}
\end{align}
which are expressed in terms of the Skyrme interaction parameters in the isospin-representation:
\begin{align}
	&C_0^{\rho,0} = \frac{3}{8}t_0 \nonumber \\
	&C_1^{\rho,0} = -\frac{1}{4}t_0\Bigg(\frac{1}{2}+x_0\Bigg) \nonumber \\
	&C_0^{\rho,\alpha} = \frac{3}{48}t_3 \nonumber \\
	&C_1^{\rho,\alpha} =  -\frac{1}{24}t_3\Bigg(\frac{1}{2}+x_3\Bigg) \nonumber \\
	&C^{\Delta\rho}_0 = -\frac{9}{64}t_1+\frac{1}{16}t_2\Bigg(\frac{5}{4}+x_2\Bigg)\nonumber\\
	&C^{\Delta\rho}_1 = \frac{3}{32}t_1\Bigg(\frac{1}{2}+x_1\Bigg)+\frac{1}{32}t_2\Bigg(\frac{1}{2}+x_2\Bigg)  \nonumber\\
	&C^{\tau}_0 = \frac{3}{16}t_1+\frac{1}{4}t_2\Bigg(\frac{5}{4}+x_2\Bigg) \nonumber\\
	&C^{\tau}_1 = -\frac{1}{8}t_1\Bigg(\frac{1}{2}+x_1\Bigg)+\frac{1}{8}t_2\Bigg(\frac{1}{2}+x_2\Bigg)
\end{align}
The generalized Lindhard functions $\chi_{2i}(k)$: $i=0,1,2$  depend on $q$ through the dimensionless variable $k=q/2k_F$. They are given by \cite{GarciaRecio:1992}:
\begin{align}
	\chi_{0}(k) = -\frac{k_F}{8\pi^2\big(\frac{\hbar^2}{2m^{*}}\big)}\Big[1+\frac{1}{2k}[1-k^2]\text{log}\Big|\frac{k+1}{k-1}\Big|\Big] \nonumber
\end{align}
\begin{align}
	\chi_{2}(k) = -\frac{k_F}{16\pi^2\big(\frac{\hbar^2}{2m^{*}}\big)}\Big[3+k^2+(1+k^2)\frac{1}{2k}[1-k^2]\text{log}\Big|\frac{k+1}{k-1}\Big|\Big] \nonumber
\end{align}
\begin{align}
	\chi_{4}(k) &= -\frac{k_F}{24\pi^2\big(\frac{\hbar^2}{2m^{*}}\big)} \Big[5+\frac{49}{3}k^2+k^4 + \nonumber\\
	& (1+k^2+k^4)\frac{1}{2k}[1-k^2]\text{log}\Big|\frac{k+1}{k-1}\Big|\Big]
\end{align}
The effective mass for homogeneous neutron matter with the Skyrme interaction is given by:
\begin{align}
\frac{\hbar^2}{2m^*(\bvec{r})} = \frac{\hbar^2}{2m}+\frac{1}{8}\big[t_1(1-x_1)+3t_2(1+x_2)\big]\rho_0
\end{align}.

\subsection{CSR applied to Skyrme}

We can easily apply the compressibility sum rule to analytically compute the $q=0$ response for a Skyrme interaction in the TL. For the homogeneous system we have:
\begin{align}
\frac{1}{\chi(0)}=-\frac{\partial^2(\rho_0\bar{E}(\rho_0))}{\partial \rho_0^2}\Bigg|_{T=0}=-\frac{d^2\mathcal{H}(\rho_0)}{d \rho_0^2}
\end{align}
and using the analytic expression for the EDF in the TL:
\begin{align}
\mathcal{H}=&\frac{\hbar^2}{2m}\tau+(C_0^{\rho,0}+C_1^{\rho,0})\rho_0^2\nonumber\\
&+(C_0^{\rho,\alpha}+C_1^{\rho,\alpha})\rho_0^{2+\alpha}+(C_0^{\tau}+C_1^{\tau})\rho_0\tau\nonumber\\
&=\frac{\hbar^2}{2m}\frac{3}{5}(3\pi^2)^{2/3}\rho_0^{5/3}+(C_0^{\rho,0}+C_1^{\rho,0})\rho_0^2\nonumber\\
&+(C_0^{\rho,\alpha}+C_1^{\rho,\alpha})\rho_0^{2+\alpha}+(C_0^{\tau}+C_1^{\tau})\frac{3}{5}(3\pi^2)^{2/3}\rho_0^{8/3},
\end{align}

the CSR yields:

\begin{align}
-\frac{\chi(0)}{\rho_0}&=\frac{1}{\rho_0}\Bigg[\frac{d^2\mathcal{H}(\rho_0)}{d \rho_0^2}\Bigg]^{-1}\nonumber\\
&=\Big[\frac{\hbar^2}{2m}\frac{2}{3}(3\pi^2)^{2/3}\rho_0^{2/3}+2(C_0^{\rho,0}+C_1^{\rho,0})\rho_0\nonumber\\
&+(2+\alpha)(1+\alpha)(C_0^{\rho,\alpha}+C_1^{\rho,\alpha})\rho_0^{1+\alpha}\nonumber\\
&+(C_0^{\tau}+C_1^{\tau})\frac{8}{3}(3\pi^2)^{2/3}\rho_0^{5/3}\Big]^{-1}
\end{align}

For calculations involving a finite number of particles, the second derivative in the CSR can be taken via a finite difference approach:
\begin{align}
&-\frac{1}{\chi(0)}\approx\frac{1}{h^2}\times\nonumber\\
&\Bigg((\rho_0+h)\bar{E}(\rho_0+h)-2\rho_0\bar{E}(\rho_0)+(\rho_0-h)\bar{E}(\rho_0-h)\Bigg)
\end{align}
We applied this approximation using a density spacing of $h=0.01\, {\rm fm}^{-3}$ for several particle numbers and found an excellent match with the analytic $\chi(0)$. 
(We also tested for $h$ values around $0.01\, {\rm fm}^{-3}$ and found that the CSR estimate had essentially flattened out/converged.) 
In all cases the estimate given by the finite difference gave very similar results. Thus we have that $66$ particles yields basically the same thing as $8250$ particles
(which, in turn, is almost identical to the analytic TL value). 
This is qualitatively simple to understand: the CSR comes from the homogeneous EOS:
the finite-size effects in the homogeneous case are dramatically smaller than the
finite-size effects of the static-response problem (which are studied
in the rest of the present paper).

\end{document}